*Article*

# Power Density and Thermochemical Properties of Hydrogen Magnetohydrodynamic (H2MHD) Generators at Different Pressures, Seed Types, Seed Levels, and Oxidizers

**Osama A. Marzouk**

College of Engineering, University of Buraimi, Al Buraimi 512, Sultanate of Oman; osama.m@uob.edu.om

**Abstract:** Hydrogen and some of its derivatives (such as e-methanol, e-methane, and e-ammonia) are promising energy carriers that have the potential to replace conventional fuels, thereby eliminating their harmful environmental impacts. An innovative use of hydrogen as a zero-emission fuel is forming weakly ionized plasma by seeding the combustion products of hydrogen with a small amount of an alkali metal vapor (cesium or potassium). This formed plasma can be used as a working fluid in supersonic open-cycle magnetohydrodynamic (OCMHD) power generators. In these OCMHD generators, direct-current (DC) electricity is generated straightforwardly without rotary turbogenerators. In the current study, we quantitatively and qualitatively explore the levels of electric conductivity and the resultant volumetric electric output power density in a typical OCMHD supersonic channel, where thermal equilibrium plasma is accelerated at a Mach number of two (Mach 2) while being subject to a strong applied magnetic field (applied magnetic-field flux density) of five teslas (5 T), and a temperature of 2300 K (2026.85 °C). We varied the total pressure of the pre-ionization seeded gas mixture between 1/16 atm and 16 atm. We also varied the seed level between 0.0625% and 16% (pre-ionization mole fraction). We also varied the seed type between cesium and potassium. We also varied the oxidizer type between air (oxygen–nitrogen mixture, 21–79% by mole) and pure oxygen. Our results suggest that the ideal power density can reach exceptional levels beyond 1000 MW/m$^3$ (or 1 kW/cm$^3$) provided that the total absolute pressure can be reduced to about 0.1 atm only and cesium is used for seeding rather than potassium. Under atmospheric air–hydrogen combustion (1 atm total absolute pressure) and 1% mole fraction of seed alkali metal vapor, the theoretical volumetric power density is 410.828 MW/m$^3$ in the case of cesium and 104.486 MW/m$^3$ in the case of potassium. The power density can be enhanced using any of the following techniques: (1) reducing the total pressure, (2) using cesium instead of potassium for seeding, and (3) using air instead of oxygen as an oxidizer (if the temperature is unchanged). A seed level between 1% and 4% (pre-ionization mole fraction) is recommended. Much lower or much higher seed levels may harm the OCMHD performance. The seed level that maximizes the electric power is not necessarily the same seed level that maximizes the electric conductivity, and this is due to additional thermochemical changes caused by the additive seed. For example, in the case of potassium seeding and air combustion, the electric conductivity is maximized with about 6% seed mole fraction, while the output power is maximized at a lower potassium level of about 5%. We also present a comprehensive set of computed thermochemical properties of the seeded combustion gases, such as the molecular weight and the speed of sound.

# 1. Introduction

## *1.1. Background*

Hydrogen is the lightest element in the periodic table, but it is the most plentiful element in the universe [1,2] (oxygen is the most abundant element in the earth's crust by mass [3,4]). Hydrogen is an important building block in several chemical processes and power systems [5,6], as listed in Table 1.

**Table 1.** Example uses of hydrogen.

| Serial Number | Characteristics | References |
|---|---|---|
| 1. | Oil refinery | [7–14] |
| 2. | Fuel cell power units | [15–19] |
| 3. | Synthesizing ammonia | [20–25] |
| 4. | Electrified-type hydrogen-powered transport | [26–33] |
| 5. | Synthesizing methanol | [34–39] |
| 6. | Gas turbines powered by hydrogen or hydrogen-blended gas | [40–42] |
| 7. | Synthesizing hydrocarbon fuels | [43–52] |
| 8. | Aerospace and rocket propulsion | [53–62] |
| 9. | Reduction processes to extract a metal from its ore | [63–71] |
| 10. | Combustion-type hydrogen-powered transport | [72–74] |
| 11. | Specialized welding operations | [75–80] |
| 12. | Food industry | [81–86] |
| 13. | Cooling of turbogenerator winding | [87–90] |
| 14. | Feedstock for chemical industries | [91–94] |

Hydrogen facilitates the transition to a sustainable (green) low-emission economy [95–98], taking advantage of the absence of released carbon dioxide ($CO_2$) due to the consumption of hydrogen [99–102]. The production of green hydrogen is based on water electrolysis powered by electricity generated from renewable energy sources (such as solar energy [103–105] and wind energy [106–109]), and thus it does not involve the release of greenhouse gases (GHG) [110–112], making it an attractive way to mitigate GHG emissions without resorting to carbon capture technologies as needed in the case of blue hydrogen [113–116] that is originating from fossil fuels.

In an earlier study, we proposed a novel use for hydrogen (either green hydrogen or blue hydrogen) in open-cycle magnetohydrodynamic power generators [117], in which direct power extraction (DPE) is realized with the aid of the Lorentz force. In this concept, hot gases formed as combustion products of hydrogen are seeded with a small amount of an alkali metal (either cesium "Cs" or potassium "K") [118–120], which allows for an appreciable level of thermal ionization leading to a suitable electric conductivity of the weakly ionized plasma gas, which is accelerated through a linear channel [121–124]. The high-speed plasma loses part of its energy in reaction to an applied magnetic field, where an induced electric field allows collection of an induced electric current to power an external electric load [125,126]. This concept of hydrogen-powered magnetohydrodynamic power generation, which we refer to here as H2MHD, can be combined with a classical power cycle [127–129] for better utilization of the input heat released from the zero-emission hydrogen combustion through a dual-cycle power plant, comprising a higher-temperature open-cycle MHD generator and either a lower-temperature closed-cycle Rankine system or a lower-temperature open-cycle Brayton system [130–132]. A triple combined cycle that incorporates the open-cycle MHD generator as a topping layer (high temperature) with an intermediate layer having an open-cycle Brayton gas turbine (intermediate temperature) and then a layer of a closed-cycle steam turbine (low temperature) is also possible, to recover part of the rejected heat containing in the still-hot plasma gas leaving the MHD generator [133–135]. Introducing an even lower-temperature refrigerant-based organic Rankine cycle [136–138] as an additional terminal downstream level (very low temperature) was proposed, yielding a quadruple combined cycle [139].

Figure 1 depicts how the proposed hydrogen magnetohydrodynamic (H2MHD) topping cycle operates. The plasma velocity vector is designated by the symbol ($\vec{u}$), the applied magnetic field vector (applied magnetic-field flux density vector) is designated by the symbol ($\vec{B}$), and the electric current-density vector is designated by the symbol ($\vec{J}$). Hydrogen is combusted with an oxidizer, which is either air (in the case of the conventional air-combustion mode) or oxygen (in the case of the intensified oxy-combustion mode). Oxy-combustion (the combustion of HHO or Brown's gas) offers the advantage of elevating the combustion temperature due to the absence of diluent nitrogen [140–142]. A small amount of solid alkali salt, such as potassium carbonate ($K_2CO_3$) [143] is added in the form of an aqueous solution that is introduced as a fine spray by atomization, to serve as the source of the vapor alkali metal after combustion [144,145]. Either potassium element or cesium element is readily vaporized well below the combustion temperature. Potassium vaporizes at 760 °C only [146], while cesium vaporizes at a lower boiling point of 671 °C [147]. The hot-seeded hydrogen-combustion products are accelerated to supersonic speeds using a de Laval nozzle (convergent–divergent nozzle or CD nozzle) [148–150]. The term "channel" or "linear channel" here refers to the portion of the divergent section of the convergent–divergent nozzle where an external magnetic field is applied perpendicular to the plasma's direction of bulk motion, and an induced electric emf (electromotive force) is produced; thus, an electric current-density can be collected through electrodes (positive cathode and negative anode) to be delivered to an external load. Like solar photovoltaic (PV) modules [151,152] and thermoelectric generators (TEG) [153,154], MHD generators operate on the principle of direct power conversion (from input energy to electricity in a single step), without an intermediate turbogenerator [155,156], and without reciprocating or rotary parts [157,158]. This is a remarkable feature in MHD generators.

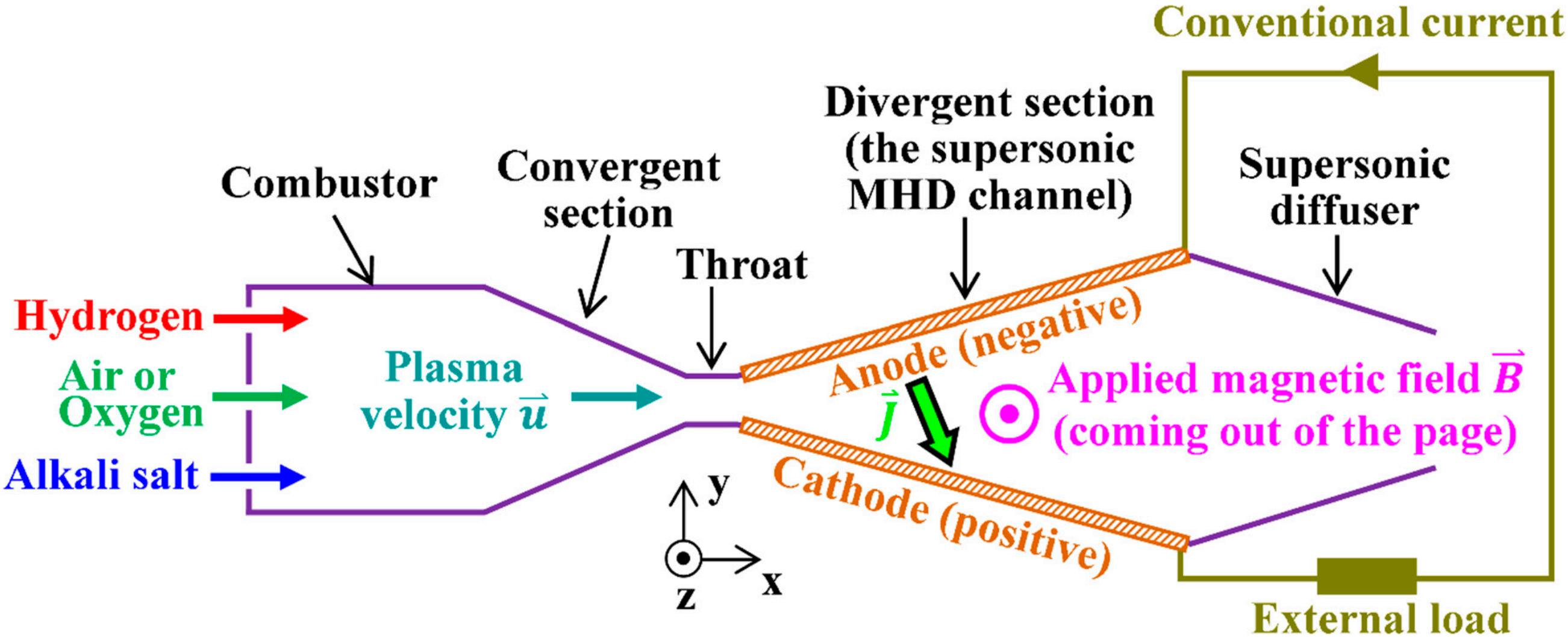

**Figure 1.** The hydrogen magnetohydrodynamic generator power system.

The power density (or volumetric power density) is a performance metric for power generation units, describing the power output per unit volume occupied. This metric quantifies the effective utilization of space. For MHD generators, the intensive high temperature and high rate of power extraction from the plasma gives this technology a unique favorable characteristic when compared to other power technologies. For example, the Sakhalin MHD pulsed power generator experimentally demonstrated a temporary power density of about 90.7 $MW/m^3$ (computed as an electric power output of 510 MW divided by an MHD linear channel volume of 5.625 $m^3$) [159–161]. Another shock-tunnel MHD generator experimentally demonstrated a high power density of 140 $MW/m^3$ [162]. These power densities are much higher than typical values of automotive engines (around 15 $MW/m^3$, the volume here refers to the displacement volume [163,164]), permanent magnet (PM) synchronous motors (SM) (around 7.5 $MW/m^3$ [165]), and triboelectric nanogenerator (TENG) energy-harvesting units (up to about 5 $MW/m^3$ [166]), and microbial fuel cells (MFCs) (between $1 \times 10^{-6}$ $MW/m^3$ and $5 \times 10^{-5}$ $MW/m^3$ [167–169]). Numerical two-dimensional modeling of a hypothetical MHD channel with methane ($CH_4$) combustion suggested a lower power density of 17.9 $MW/m^3$ (computed through dividing an electric power of 31.4 MW by a volume of 1.75 $m^3$) [170]. This is not remarkably high, but it is still encouraging.

There is another metric for expressing the effective utilization of a power system, which is the specific power of gravimetric power density (W/kg). However, this is more challenging to estimate compared to the volumetric power density; estimating the specific power requires more detailed and precise knowledge about the particular power system and its components, including the minor auxiliary ones if they largely affect the weight [171,172]. The specific power (or the power-to-weight ratio [173,174]) is of particular importance in aeronautical and space propulsion systems [175–177], which are outside the scope of this study that is concerned with terrestrial power systems. Thus, the volumetric power density is adopted here. The specific energy or the gravimetric energy density metric (Wh/kg) [178,179], and the energy density or volumetric energy density metric (Wh/L) [180,181] pertain more to energy storage units (such as batteries), and thus are not used in our study.

### *1.2. Goal of the Study*

Because the proposed H2MHD concept is a new technology for utilizing hydrogen, there is a need for extensive investigations regarding the technological feasibility of this technology. This study aims to partly fill this huge knowledge gap through estimating the volumetric power density of an ideal H2MHD channel, at a representative value for the applied magnetic field (the applied magnetic-field flux density), a reasonable supersonic Mach number (the ratio of the plasma speed to the speed of sound), and a suitable temperature representative of hydrogen combustion.

The performance of the volumetric power density is estimated for a wide range of operational conditions, with different total pressures, different levels of seeding, two types of the seed metal, and two types of the oxidizer.

The results of this study provide several insights into the use of OCMHD in general, rather than H2MHD solely because some of the generic findings reported here can actually be applicable to other OCMHD systems with fuels other than hydrogen, including coal and fuels derived from renewable biomass [182–185]. Thus, this study is considered useful for MHD power generators in general, and for hydrogen use in these power generators in particular.

## 2. Research Method

### 2.1. Assumptions

The current study is centered on zero-dimensional computational modeling (point modeling or system-level modeling) of a hypothetical supersonic hydrogen magneto-hydrodynamic channel. Therefore, no exact geometry is defined; instead, the analysis assumes uniform local properties everywhere. Despite losing boundary effects and temporal transience [186–191] that can be captured using computational fluid dynamics (CFD) methods [192–196], our simplifying assumption is important to allow having a closed-form expression for the power density, which is conveniently applied to estimate theoretical upper bounds. In addition, our assumption is useful in generalizing the outcomes of this study, making them broadly valid, rather than being limited to a specific H2MHD channel.

In addition, due to the large number of variables involved in the H2MHD problem, we need to fix some variables at reasonable values while changing others to explore their impact. Therefore, three quantities are fixed in the current study, and these fixed quantities are (1) the temperature, (2) the applied magnetic-field flux density, and (3) the Mach number. These fixed parameters are set to the values listed in Table 2. The significance of these fixed parameters lies also in the fact they are the same values assumed in our previous work mentioned earlier in which we proposed the H2MHD concept (but the term "M2HHD" is coined here for the first time).

**Table 2.** Three fixed parameters in the current study.

| Fixed Parameter | Value | References |
|---|---|---|
| Temperature | 2300 K = 2026.85 °C | [197–199] |
| Magnetic-field flux density | 5 T (5 teslas) = 5000 G (50,000 gausses) | [200–203] |
| Mach number | 2 | [204–206] |

### 2.2. Mathematical Equations

For a plasma speed ($u$), an applied magnetic-field flux density magnitude ($B$), and an electric conductivity ($\sigma$), the theoretical upper limit for the volumetric electric power density ($P_V$) that can be extracted from the moving plasma gas is [207].

$$P_V = \frac{1}{4}\sigma\, u^2 B^2 \tag{1}$$

This power density is restricted to ideal conditions, where the external load is matched (optimized) such that its electric resistance is equal to the equivalent internal electric resistance of the MHD generator [208–211].

The plasma speed ($u$) can be expressed in terms of the speed of sound ($a$) and the Mach number (M) as [212].

$$u = \mathrm{M}\, a \tag{2}$$

Using Equation (2) in Equation (1) allows for expressing the power density ($P_V$) in a different form as

$$P_V = 0.25\, \sigma\, \mathrm{M}^2\, a^2\, B^2 \tag{3}$$

The plasma electric conductivity ($\sigma$) in our study is computed according to an algorithm we described and verified in detail in a previous study [213], which is based on the robust Frost's two-conductivity approximation that takes into account the scattering of electrons by neutrals as well as the scattering of electrons by charged particles (ions and other electrons) [214], and thus interpolates the electric conductivity for the partially ionized plasma in OCMHD channels between the two extreme values for which an exact solution is known (namely, the extreme case of very weakly ionized plasma, and the extreme case of fully ionized plasma) [215–217]. In our algorithm, the plasma's electric conductivity depends on the total static pressure ($p$) (the term "total" here means the sum of partial pressures of involved gaseous species, not the stagnation zero-speed pressure, and not the gauge pressure [218]), the absolute temperature ($T$), and the pre-ionization mole fractions ($X_i$, where $i$ is an index referring to each involved gaseous species in the gaseous mixture formed after adding the seed alkali metal vapor to the hydrogen combustion products). Since the temperature is fixed in the current study at 2300 K, practically, the electric conductivity depends only on the total static pressure ($p$) and the mole fractions ($X_i$). Therefore, we can symbolically represent the electric conductivity ($\sigma$) as ($\sigma(p, X_i)$) to emphasize its functional dependence on these plasma properties.

The speed of sound in the pre-ionization plasma ($a$) is modeled here using the ideal gas approximation, making it a function of the absolute temperature ($T$) and the mole fractions ($X_i$) only [219–221]. This speed of sound can be expressed as

$$a = \sqrt{\frac{C_{p,mix}}{C_{p,mix} - R_{mix}} R_{mix}\ T} \tag{4}$$

where ($C_{p,mix}$) is the specific heat capacity at constant pressure of the gas mixture after seeding but before ionization, and ($R_{mix}$) is the specific gas constant for this gas mixture. The pre-ionization mixture's specific heat capacity ($C_{p,mix}$) is computed from the specific heat capacities of the individual gaseous species ($C_{p,i}$ for the species with the index $i$) that make up the pre-ionization gas mixture, and these individual specific heat capacities are described according to the updated nine-coefficient edition of NASA (United States National Aeronautics and Space Administration) nonlinear fitting functions for computing thermodynamic properties [222–225]. Therefore, we can analytically express the speed of sound ($a$) as ($a(T, X_i)$) to emphasize its functional dependence on the plasma properties ($T$) and ($X_i$).

The nondimensional quantity ($C_{p,mix}/(C_{p,mix} - R_{mix})$) in Equation (4) is called the specific heat ratio or the adiabatic index, and here it is assigned the symbol ($\gamma_{mix}$). Therefore, the previous equation for the speed of sound ($a$) can be rewritten as [226]

$$a = \sqrt{\gamma_{mix}\ R_{mix}\ T} \tag{5}$$

The gas constant ($R_i$) for each gaseous species having the index $i$ is obtained using

$$R_i = \frac{\overline{R}}{M_i} \tag{6}$$

where ($\overline{R}$) is the universal (molar) gas constant, and ($M_i$) is the molecular weight of the species with an index ($i$). In the current study, we have [227–229]

$$\overline{R} = 8314.462618 \frac{\text{J}}{\text{kmol.K}} \tag{7}$$

The molecular weights of the gaseous species that appear in the current study are listed in Table 3. These were taken values from the NIST (United States National Institute of Standards and Technology) individual gaseous species Chemistry WebBook database [230–233].

**Table 3.** Molecular weights used in the current study.

| Gaseous Species | Molecular Weight [kg/kmol] | Reference |
|---|---|---|
| Water vapor ($H_2O$) | 18.0153 | [234] |
| Nitrogen ($N_2$) | 28.0134 | [235] |
| Cesium vapor (Cs) | 132.9054519 | [236] |
| Potassium vapor (K) | 39.0983 | [237] |

The molecular weight of the gas mixture ($M_{mix}$) is computed as [238]

$$M_{mix} = \sum_{i=1}^{n_s} M_i\, X_i \tag{8}$$

where ($n_s$) is the number of gaseous species in the pre-ionization gas mixture. We have ($n_s = 2$) when the oxidizer is pure oxygen (oxy-hydrogen combustion), while ($n_s = 3$) when the oxidizer is air (air–hydrogen combustion). In the case of oxy-combustion, the pre-ionization mixture is treated here as a mixture of water vapor ($H_2O$) and the vaporized alkali metal (either Cs or K). In the case of air–hydrogen combustion, the pre-ionization mixture is treated here as a mixture of water vapor ($H_2O$), excess nitrogen ($N_2$), and the vaporized alkali metal (either Cs or K).

After obtaining ($M_{mix}$), the specific gas constant of the pre-ionization gas mixture ($R_{mix}$) can be computed as

$$R_{mix} = \frac{\overline{R}}{M_{mix}} \tag{9}$$

Then, the mass fraction of each individual species ($Y_i$) can be computed as [239]

$$Y_i = \frac{X_i\, M_i}{M_{mix}} \tag{10}$$

Then, the specific heat capacity of the mixture is computed from the specific heat capacities of the individual species through mass-weighted averaging as

$$C_{p,mix} = \sum_{i=1}^{n_s} C_{p,i}\, Y_i \tag{11}$$

The NASA nonlinear fitting function for estimating ($C_{p,i}$) has the following general structure, with seven terms:

$$C_{p,i} = R_i \left( b_{1,i}\, T^{-2} + b_{2,i}\, T^{-1} + b_{3,i} + b_{4,i}\, T + b_{5,i}\, T^2 + b_{6,i}\, T^3 + b_{7,i}\, T^4 \right) \tag{12}$$

It should be noted that the fitting coefficients ($b_{1,i}$ to $b_{7,i}$) in this equation correspond to the higher temperature range (from 1000 K to 6000 K); there is another set of fitting coefficients for a lower-temperature range (from 200 K to 1000 K), but these are irrelevant in our study (too cold to be of interest) (Table 4).

The combustion of hydrogen with air is approximated as a complete stoichiometric reaction of molecular hydrogen ($H_2$) with a mixture having mole fractions of 79% for molecular nitrogen ($N_2$) and 21% for molecular oxygen ($O_2$) [240–242]. Therefore, the combustion products are idealized as having only water vapor ($H_2O$) and excess non-reacting molecular nitrogen ($N_2$) according to the following reaction equation [243,244]:

$$2H_2 + O_2 + 3.762N_2 \rightarrow 2H_2O + 3.762N_2 \tag{13}$$

**Table 4.** Fitting coefficients of the specific heat capacities at constant pressure.

| Coefficient for ($C_{p,i}$) | Species Index ($i$) | | | |
|---|---|---|---|---|
| | $H_2O$ | $N_2$ | Cs | K |
| $b_{1,i}$ | $1.034972096 \times 10^{6}$ | $5.877124060 \times 10^{5}$ | $6.166040900 \times 10^{6}$ | $-3.56642236 \times 10^{6}$ |
| $b_{2,i}$ | $-2.412698562 \times 10^{3}$ | $-2.239249073 \times 10^{3}$ | $-1.896175522 \times 10^{4}$ | $1.085289825 \times 10^{4}$ |
| $b_{3,i}$ | $4.646110780$ | $6.066949220$ | $2.483229903 \times 10^{1}$ | $-1.054134898 \times 10^{1}$ |
| $b_{4,i}$ | $2.291998307 \times 10^{-3}$ | $-6.139685500 \times 10^{-4}$ | $-1.251977234 \times 10^{-2}$ | $8.009801350 \times 10^{-3}$ |
| $b_{5,i}$ | $-6.836830480 \times 10^{-7}$ | $1.491806679 \times 10^{-7}$ | $3.309017390 \times 10^{-6}$ | $-2.696681041 \times 10^{-6}$ |
| $b_{6,i}$ | $9.426468930 \times 10^{-11}$ | $-1.923105485 \times 10^{-11}$ | $-3.354012020 \times 10^{-10}$ | $4.715294150 \times 10^{-10}$ |
| $b_{7,i}$ | $-4.822380530 \times 10^{-15}$ | $1.061954386 \times 10^{-15}$ | $9.626500908 \times 10^{-15}$ | $-2.976897350 \times 10^{-14}$ |

Vapor seed metal (either cesium or potassium) is then added to the hydrogen combustion products with a predetermined target mole fraction. This seeded mixture is then analyzed under thermal equilibrium to obtain the electrochemical properties of the post-ionization plasma (such as its electric conductivity and mole fractions of ions).

The combustion of hydrogen with oxygen is approximated as a complete stoichiometric reaction of molecular hydrogen ($H_2$) with molecular oxygen ($O_2$), resulting in pure water vapor ($H_2O$) as the single combustion product. This is described by the following reaction equation:

$$2H_2 + O_2 \rightarrow 2H_2O \tag{14}$$

Vapor seed metal (either cesium or potassium) is then added to this combustion product water vapor with a predetermined target mole fraction. This seeded mixture is then analyzed under thermal equilibrium to obtain the electrochemical properties of the post-ionization plasma.

We conclude the current mathematical description of the modeling methodology by presenting a customized version of the expression governing the ideal volumetric power density in Equation (3), by setting the Mach number ($\mathrm{M} = 2$) and the magnitude of the applied magnetic-field flux density ($B = 5$ T). This gives the following special expression pertaining to our study:

$$P_V \left[\frac{\mathrm{W}}{\mathrm{m}^3}\right] = 25\,\sigma \left[\frac{\mathrm{S}}{\mathrm{m}}\right] \left(a \left[\frac{\mathrm{m}}{\mathrm{s}}\right]\right)^2 \tag{15}$$

To conveniently express the values of the power density ($P_V$), the larger unit of ($\mathrm{MW/m^3}$) is adopted rather than the basic SI unit of ($\mathrm{W/m^3}$), thus the previous equation is modified to

$$P_V \left[\frac{\mathrm{MW}}{\mathrm{m}^3}\right] = \frac{25}{10^6}\,\sigma \left[\frac{\mathrm{S}}{\mathrm{m}}\right] \left(a \left[\frac{\mathrm{m}}{\mathrm{s}}\right]\right)^2 \tag{16}$$

While the speed of sound ($a$) in the above equation should correspond to the post-ionization plasma condition (taking into account the small changes in the gaseous mixture because of the partial ionization of some alkali seed atoms), we use the pre-ionization value that corresponds to an all-neutral (no ions, no electrons) seeded gas mixture. This establishes consistency with the reported total pressure values as a controllable variable, where these pressures correspond to the pre-ionization condition. This matter is practically negligible because only a small portion of the seeded alkali metal can be thermally ionized at the fixed temperature of 2300 K, causing the mole fraction of either the alkali ions or the free electrons to be below 0.1% in all cases we cover, and even much less than this in many of the analyzed cases.

### 2.3. Varied Conditions

In Table 5, we list the quantities to be varied in the current study while estimating the ideal volumetric power density ($P_V$) for a magnetohydrodynamic (MHD) generator channel, as well as the values assigned to these varied quantities. These quantities are listed below:

**Table 5.** Varied quantities in the current study.

| Counter | Varied Quantity | | | |
|---|---|---|---|---|
| | $p$ [atm] | $X_{seed}$ [%] | Seed Metal | Oxidizer |
| 1. | 0.0625 | 0.0625%, (1/16)% | Cesium (Cs) | Air (79% $N_2$, 21% $O_2$) |
| 2. | 0.125 | 0.25% | Potassium (K) | Oxygen (100% $O_2$) |
| 3. | 0.25 | 1% | | |
| 4. | 0.5 | 4% | | |
| 5. | 1 | 16% | | |
| 6. | 2 | | | |
| 7. | 4 | | | |
| 8. | 8 | | | |
| 9. | 16 | | | |

(1) The total absolute pressure ($p$) after the vaporous seed alkali metal is mixed with the hydrogen combustion gases. The unit is the standard atmosphere (1 atm = 101,325 Pa = 1.01325 bar). Nine values are assigned to this continuous variable, sampling it between a very low value of 1/16 atm (6332.812 Pa or 6.332812 kPa) to a high value of 16 atm (1.6212 MPa or 1621.2 kPa). These values allow for exploring the impact of this variable over a wide range. The nine selected pressures form a geometric sequence with a multiplicative progression constant of two.

(2) The mole fraction ($X_{seed}$) of the vaporous seed alkali metal after it is mixed with the hydrogen combustion gases. This fraction is expressed as a percentage. Five values are assigned to this continuous variable, sampling it between a very low value of 0.0625% or (1/16)% to a high value of 16%. These values allow for exploring the impact of this variable over a wide range. The five selected seed mole fractions form a geometric sequence with a multiplicative progression constant of four.

(3) The type of the vaporous seed alkali metal to be added to the hydrogen combustion gases. This is a binary variable (this means it can take one of two values only). The seed type here can be either cesium (Cs) or potassium (K). These two particular chemical elements are chosen here as the possible seed metal that serves as the source of electrons due to their easy thermal ionization compared to the combustion products gases ($H_2O$ or $N_2$). The ionization energy of cesium (Cs, atomic number 55; located in the alkali metal group IA or group 1 and period 6 in the periodic table of chemical elements) is 3.893 eV/particle (375.6 kJ/mol), and this is the lowest value among all chemical elements in the periodic table [245–247]. We point out here that 1 eV/particle (electronvolt per particle) is equivalent to 96.49 kJ/mol (kilojoule per mole) [248–250]. The ionization energy of potassium (K, atomic number 19; located in the alkali metal group IA or group 1 and period 4 in the periodic table of chemical elements) is 4.34 eV/particle (419 kJ/mol), and this is the fourth lowest value among all chemical elements in the periodic table [251–253]. Although francium (Fr, atomic number 87; located in the alkali metal group IA or group 1 and period 7 in the periodic table) has a lower ionization energy of (3.9 eV/particle or 376 kJ/mol) than potassium [254,255], francium is radioactive and very unstable (has no practical uses) due to its small half-life of only 22 min [256–258]. Although rubidium (Rb, atomic number 37; located in the alkali metal group IA or group 1 and period 5 in the periodic table) has a lower ionization energy of (4.177 eV/particle or 403.0 kJ/mol) than potassium [259,260], rubidium is a rare element, and its use is typically limited to scientific research [261–263]. The ionization energy of molecular nitrogen ($N_2$) is very large, being 15.6 eV/particle (1505 kJ/mol) approximately [264,265]. The ionization energy of atomic nitrogen (N) is also large, being 14.5 eV/particle (1399 kJ/mol) approximately [266,267]. The dissociation energy of molecular nitrogen ($N_2$) into two atoms is also much lower (9.76 eV/mole or 942 kJ/mol) [268,269], but it is still higher than the ionization energy of either potassium or cesium. The ionization energy of the water molecules ($H_2O$) is very large, being 12.6 eV/particle (1216 kJ/mol) approximately [270]. The dissociation energy of the water molecules ($H_2O$) into a hydrogen atom (H) and a hydroxyl radical (OH) radical is approximately 5.16 eV/particle (498 kJ/mol) [271], making this dissociation less probable than the ionization of either potassium or cesium.

(4) The type of oxidizer to be used in the combustion of hydrogen. This is a binary variable, and it can be either air or pure oxygen ($O_2$).

Therefore, there is a total of 180 conditions (data points) to be computationally represented in the current study (because there are nine pressures, five seed mole fractions, two seed types, and two oxidizer types).

### 2.4. Power Density Criterion

In our study, it is sufficient to estimate the volumetric electric power density output under the various operational conditions for the H2MHD (hydrogen magnetohydrodynamic) concept, as this is its primary goal and major contribution. However, the value of our study is boosted when a criterion for this power density is also recommended, to help in judging the feasibility of the H2MHD concept and the preferred set of conditions suitable for it.

To justify such a criterion (a power density threshold), we consider that a mid-size power plant has a power capacity on the order of 600 MW [272–275]. If an H2MHD power density of 50 MW/m$^3$ is possible, then this 600 MW power capacity can be accomplished with a compact space of only 12 m$^3$ in the case of a properly operated H2MHD generator. This small volume can be formed, for example, as a divergent channel with a width of 1 m, a length of 6 m, and a mean height of 2 m [276–279]. To account for the assumptions made in the current study, and the losses ignored in our system-level approach, we set our threshold to twice this value, making it 100 MW/m$^3$. This means that 100 MW/m$^3$ in our analysis may actually be reduced to half of this value in reality. To show the gigantic amount of power delineated by this criterion, it is helpful to compare this threshold (100 MW/m$^3$) with the typical area power density of photovoltaic (PV) solar panels under good sunshine, which is only about 200 W/m$^2$ (0.0002 MW/m$^2$) [280–282].

We emphasize that this suggested criterion of 100 MW/m$^3$ here is not a precise value, but involves some arbitrariness. However, it is considered a reasonable rough guide for assessing the suitability of H2MHD power generation.

## 3. Results

This results section is divided into four subsections, which are mapped with the four possibilities arising from the two binary variables (namely, two alkali seed types and two oxidizer types). For each case of these four conditions, we present first the variation in the plasma electric conductivity ($\sigma$) after thermal equilibrium as a function of the total pre-ionization absolute pressure ($p$) with the mole fraction of the neutral alkali seed metal ($X_{seed}$) being a parameter. Thus, the variation in ($\sigma$) is visualized versus ($p$) as five curves (one curve per $X_{seed}$), and this visualization is presented four times (for two seed types and two oxidizer types). Similarly, the estimated ideal volumetric power density ($P_V$) is visualized as a function of ($p$) four times in the four coming subsections, at five values of ($X_{seed}$) in each ($P_V$) plot.

To aid in interpreting the electric conductivity, it is voluntarily expressed as a multiple of typical seawater electric conductivity, which is assigned here a value of 5 S/m [283–285]. The secondary (right) vertical axis in the electric conductivity ($\sigma$) plots is utilized for this purpose, while the primary (left) vertical axis is utilized for expressing the absolute (unscaled) value of ($\sigma$), expressed in the SI unit of S/m.

Regarding the pre-ionization speed of sound ($a$), and thus the plasma speed ($u$), because ($a$) is independent of the total pressure, it does not need to be visualized as a dependent variable of ($p$). Instead, we only tabulate its values for the five values of ($X_{seed}$) at the beginning of each of the four forthcoming subsections, along with other thermochemical properties that are also independent of the total absolute pressure (such as the pre-ionization mixture's molecular weight).

In addition to the visualized curves, we also list the numerical values for the variations in ($\sigma$) and ($P_V$). This is useful for the sake of precision and for subsequent use of these values by interested readers in external quantitative analysis or comparisons.

For effective contrast and use of the plots, the ranges of the axes are retained the same regardless of the actual range of the data being visualized. This allows for quick identification of the relative size of the data visualized in different figures, which correspond to different operational cases or the H2MHD channel.

For effective identification of the individual influence of the seed type and the oxidizer type, we intentionally order the four forthcoming subsections such that in the transition from one subsection to the next, only one change is made. This change either alters the seed type or alters the oxidizer type (but not altering both of them at once, which makes it difficult to separate the influence of each alteration from the other).

### 3.1. Cesium Seed and Air Oxidizer

The first set of results corresponds to the condition of using cesium vapor as the ionizable gas, and using air as the oxidizer for the hydrogen combustion.

In Table 6, we provide our computed thermochemical characteristics for the seeded gas mixture before ionization, according to the mathematical model presented in the previous section.

**Table 6.** Pre-ionization properties of cesium-seeded air–hydrogen combustion mixture.

| Property | Different Pre-Ionization Cesium Mole Fraction ($X_{Cs}$) | | | | |
|---|---|---|---|---|---|
| | **0.0625%** | **0.25%** | **1%** | **4%** | **16%** |
| $X_{Cs}$ [%] | 0.062500% | 0.250000% | 1.000000% | 4.000000% | 16.000000% |
| $X_{H2O}$ [%] | 34.688476% | 34.623395% | 34.363068% | 33.321763% | 29.156543% |
| $X_{N2}$ [%] | 65.249024% | 65.126605% | 64.636932% | 62.678237% | 54.843457% |
| $M_{mix}$ [kg/kmol] | 24.611 | 24.814 | 25.627 | 28.878 | 41.881 |
| $R_{mix}$ [J/(kg.K)] | 337.8384 | 335.0721 | 324.4457 | 287.9214 | 198.5258 |
| $\gamma_{mix}$ [-] | 1.243991 | 1.244265 | 1.245365 | 1.249869 | 1.269668 |
| $a$ [m/s] | 983.166 | 979.241 | 964.014 | 909.773 | 761.408 |
| $u$ [m/s] | 1966.33 | 1958.48 | 1928.03 | 1819.55 | 1522.82 |

It is noticed that as the mole fraction of cesium increases, the mixture's gas constant ($R_{mix}$) decreases (because the mixture's molecular weight increases). This is justified by the heavier molecular weight of cesium (Cs) compared to either water vapor ($H_2O$) or molecular nitrogen ($N_2$). One mole of cesium (Cs) is 4.744 times heavier than one mole of molecular nitrogen ($N_2$), and 7.377 times heavier than one mole of water ($H_2O$). One of the implications of the decline in the mixture's gas constant ($R_{mix}$) is a decline in the speed of sound ($a$), and thus the plasma's bulk speed (given that it is always twice the speed of sound in the current study). However, these influences of the additive cesium remain marginal up to a mole fraction of 1%. The remarkable changes at the large mole fraction of 16% are quite theoretical only, because practically it is not expected to have such a big portion of the seed.

Despite the seed content, the mole fraction ratio ($X_{N2}/X_{H2O}$) is always 1.881, as implied by the reaction stoichiometry indicated in Equation (13), which stipulates that this ratio should be $3.762/2 = 1.881$ in the cases of air–hydrogen combustion.

Figure 2 shows how the electric conductivity of the hydrogen plasma drops monotonically and nonlinearly as the total pre-ionization absolute pressure increases. This trend is identical for the five seed levels of cesium vapor. This drop in the electric conductivity can be explained by the larger number density of neutral heavy particles (atoms and molecules) in the plasma at higher pressures, which retards the directional movement of the liberated electrons from the ionized cesium atoms due to more frequent collisions with these neutral particles [286–288].

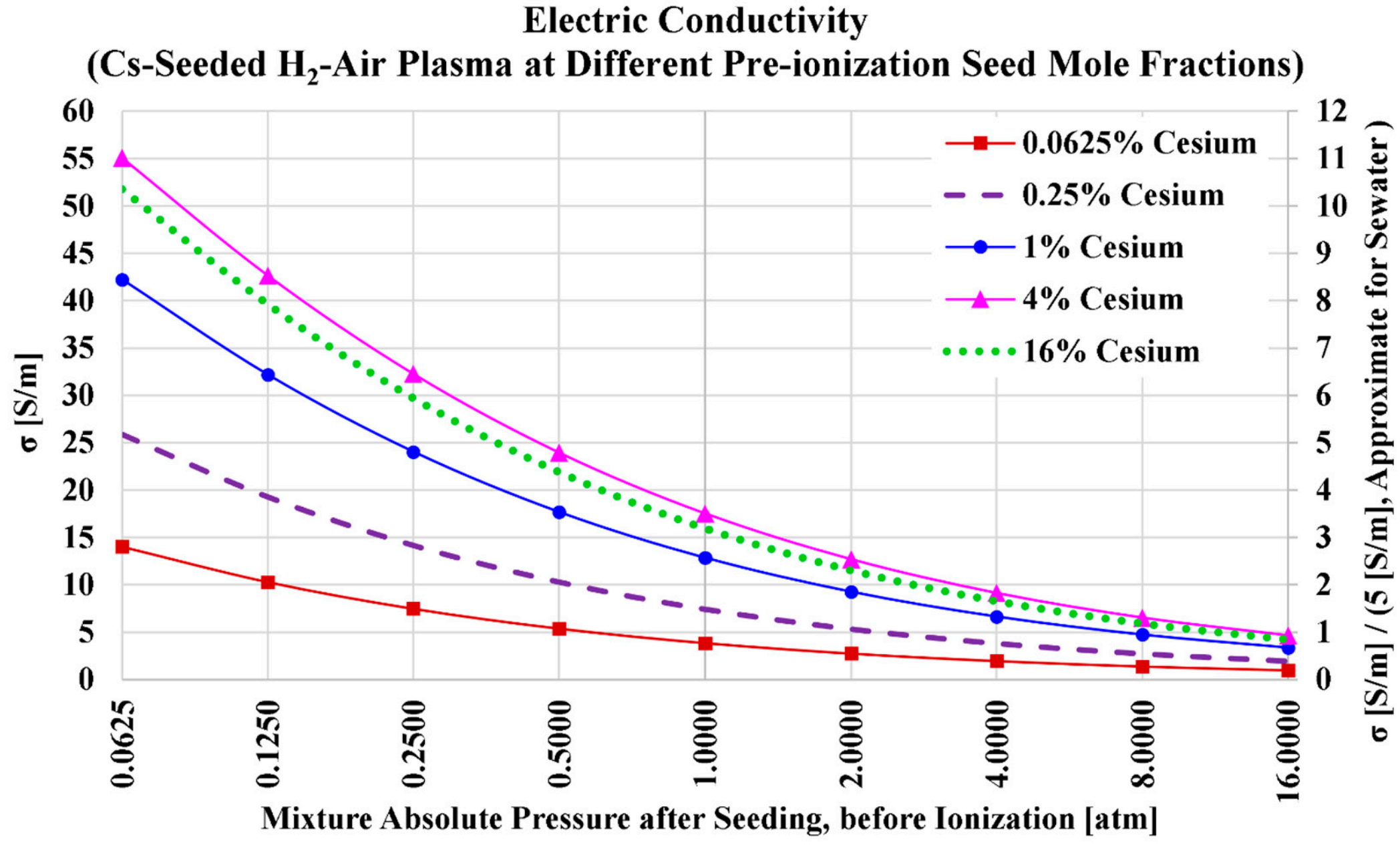

**Figure 2.** Electric conductivity in the case of cesium seed and air oxidizer.

Unlike the effect of the pressure, the effect of the cesium additive is non-monotonic. Starting from a small level of (Cs), adding more (Cs) initially improves the electric conductivity. This boost slows down and reaches a peak near a cesium mole fraction of $X_{Cs} = 6\%$, and then the electric conductivity starts to drop as more cesium vapor is added in place of water vapor and nitrogen. This peaking behavior of the electric conductivity ($\sigma$) can be attributed to the presence of two factors that impact the electric conductivity due to the free electrons produced by the ionization of a fraction of the cesium atoms. The first factor is the large number density of electrons, due to more cesium atoms; this improves the electric conductivity as more cesium is added. The other factor is the electron–ion interaction (attraction) and electron–electron interaction (repulsion) due to their electric charges. As the number of electrons (and thus the number of cesium ions) increases beyond an appreciable level, the attenuating effect on the electric conductivity caused by these interactions (Coulomb scattering) [289–291] overpowers the accentuating effect on the electric conductivity caused by the electrons' number density, and thus the electric conductivity declines after reaching a peak value. This explains why the electric conductivity with a higher cesium mole fraction of 16% is less than the electric conductivity with a lower cesium mole fraction of 4%. From the figure shown, it may be inferred qualitatively that a pre-ionization cesium mole fraction of 1% or 2% is preferred, because the weak increase in the electric conductivity beyond these levels does not seem to justify such additions of cesium content.

Table 7 lists the numerical values of the plasma electric conductivity as visualized in the previous figure. We also add to the table the electric conductivity values at a pre-ionization cesium mole fraction of $X_{Cs} = 6\%$, which approximately corresponds to the maximized electric conductivity for this case (air–hydrogen combustion plasma with cesium seed).

**Table 7.** Electric conductivity [S/m] of hydrogen plasma (cesium seed, air oxidizer).

| Pressure [atm] | Plasma Electric Conductivity [S/m] (at the Different Cesium Levels) | | | | | |
|---|---|---|---|---|---|---|
| | **0.0625%** | **0.25%** | **1%** | **4%** | **6%** | **16%** |
| 0.0625 | 14.0116 | 25.8383 | 42.2099 | 55.0610 | 56.0605 | 51.7766 |
| 0.125 | 10.2858 | 19.2426 | 32.1752 | 42.6407 | 43.4010 | 39.6267 |
| 0.25 | 7.4702 | 14.1237 | 24.0379 | 32.2535 | 32.8188 | 29.6901 |
| 0.5 | 5.3830 | 10.2551 | 17.6829 | 23.9483 | 24.3616 | 21.8837 |
| 1 | 3.8573 | 7.3878 | 12.8589 | 17.5335 | 17.8321 | 15.9340 |
| 2 | 2.7530 | 5.2923 | 9.2729 | 12.7050 | 12.9189 | 11.4995 |
| 4 | 1.9593 | 3.7762 | 6.6470 | 9.1379 | 9.2904 | 8.2470 |
| 8 | 1.3917 | 2.6869 | 4.7447 | 6.5378 | 6.6461 | 5.8882 |
| 16 | 0.9871 | 1.9081 | 3.3768 | 4.6603 | 4.7371 | 4.1912 |

Considering the five selected seed levels (0.0625%, 0.25%, 1%, 4%, and 16%), the highest obtained electric conductivity is $\sigma = 55.0610$ S/m (at $X_{Cs} = 4\%$ and $p = 0.0625$ atm), and this is about 11 times the electric conductivity of seawater. At the higher $X_{Cs} = 16\%$ (and the same total pressure of 0.0625 atm), the electric conductivity is slightly lower, with a value of 51.7766 S/m; at the lower $X_{Cs} = 1\%$ (and the same total pressure of 0.0625 atm), the electric conductivity is mildly lower, with a value of 42.2099 S/m. If we take a total pressure of 1 atm (atmospheric MHD channel) as a reference, one can see that the electric conductivity can exceed 10 S/m (twice the conductivity of seawater) with only 1% cesium level.

Using the presented values of the speed of sound (see Table 6) and the values of the electric conductivity (see Table 7) in Equation (16) gives the corresponding ideal volumetric power densities ($P_V$) for the current H2MHD case of cesium seeding and air oxidizer. The ($P_V$) results are visualized in Figure 3 and listed in Table 8. We also add to the table the power density values at a pre-ionization cesium mole fraction of $X_{Cs} = 3\%$, which approximately corresponds to the maximized power density for this case (air–hydrogen combustion plasma with cesium seed).

Like the electric conductivity, the power density declines monotonically and nonlinearly as the total pressure increases. Also, like the electric conductivity, the power density exhibits a peak when the cesium level increases, and then it starts to decrease. However, unlike the case of the electric conductivity ($\sigma$), the cesium level for maximizing the power density ($P_V$) has shifted down from being close to 6% to being close to 3%. This shift is explained by the simultaneous decline in the speed of sound as the cesium level increases. This added effect of the speed of sound (which quadratically affects the power density) favors lower cesium values for maximizing the power density.

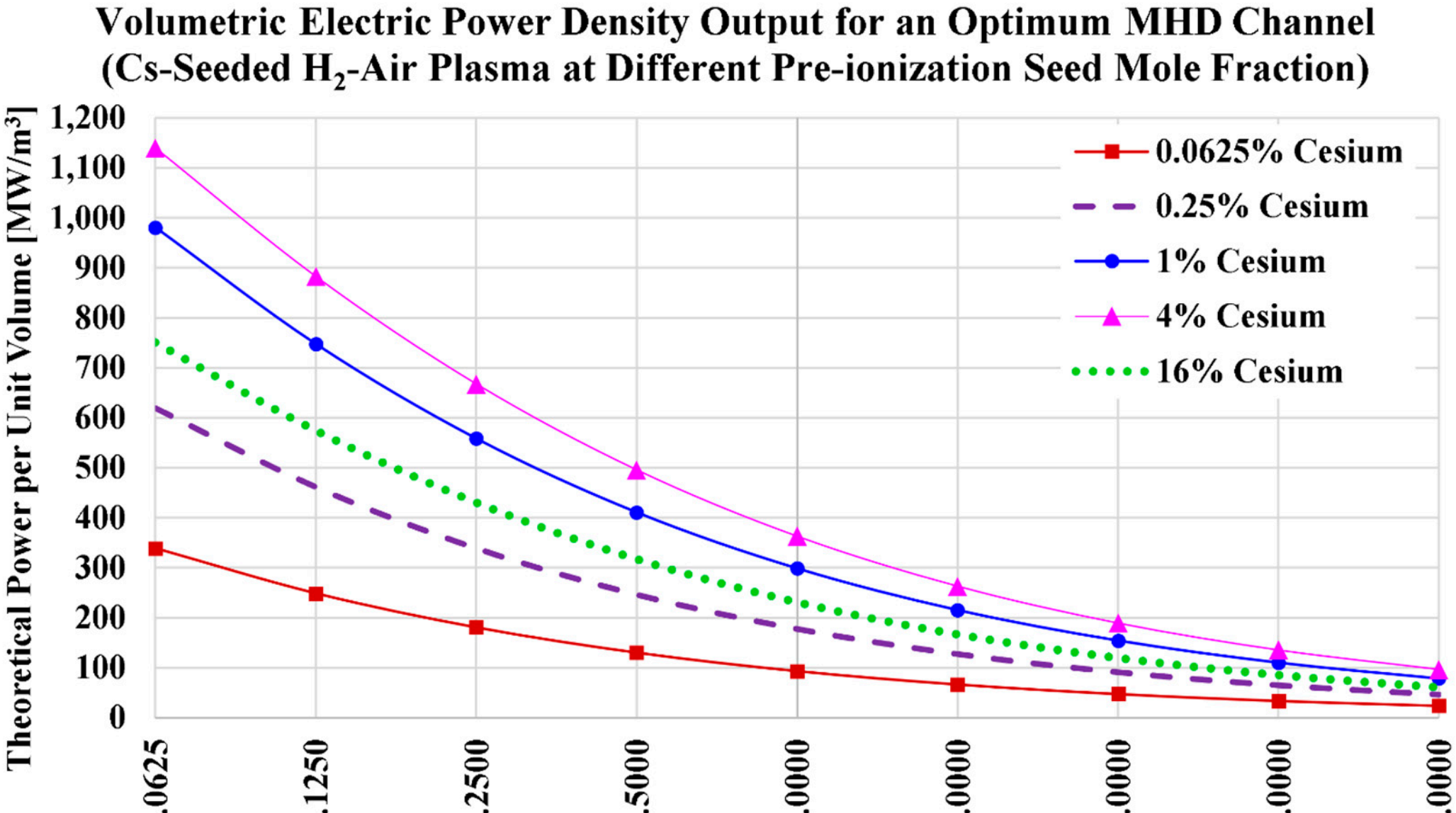

**Figure 3.** Ideal power density in the case of cesium seed and air oxidizer.

**Table 8.** Power density [MW/m$^3$] of hydrogen plasma (cesium seed and air oxidizer).

| Pressure [atm] | Volumetric Power Density [MW/m$^3$] (at the Different Cesium Levels) | | | | | |
|---|---|---|---|---|---|---|
| | **0.0625%** | **0.25%** | **1%** | **3%** | **4%** | **16%** |
| 0.0625 | 338.596 | 619.417 | 980.665 | 1146.828 | 1139.332 | 750.427 |
| 0.125 | 248.561 | 461.299 | 747.529 | 886.885 | 882.329 | 574.332 |
| 0.25 | 180.520 | 338.585 | 558.474 | 670.107 | 667.395 | 430.315 |
| 0.5 | 130.082 | 245.844 | 410.828 | 497.150 | 495.543 | 317.173 |
| 1 | 93.213 | 177.106 | 298.752 | 363.771 | 362.806 | 230.940 |
| 2 | 66.527 | 126.871 | 215.438 | 263.484 | 262.894 | 166.669 |
| 4 | 47.347 | 90.526 | 154.430 | 189.454 | 189.083 | 119.528 |
| 8 | 33.631 | 64.413 | 110.234 | 135.521 | 135.281 | 85.341 |
| 16 | 23.854 | 45.743 | 78.453 | 96.591 | 96.432 | 60.745 |

The tabulated data are helpful in quantifying potential ranges of the power density in the current case of cesium seed and air oxidizer. Considering the five selected seed levels (0.0625%, 0.25%, 1%, 4%, and 16%), the best obtained power density is 1139.332 MW/m$^3$ (at 4% cesium and 0.0625 atm), and the worst obtained power density is 23.854 MW/m$^3$ (at 0.0625% cesium and 16 atm). At atmospheric pressure, a promising power density of 362.806 MW/m$^3$ is attainable with 4% cesium, and an attractive power density of 298.752 MW/m$^3$ is attainable with only 1% cesium. Even with 0.25% cesium, a power density of 177.106 MW/m$^3$ can be achieved at atmospheric combustion. These are favorably intense power densities, which we found to require a minute amount of cesium seed (at 1 atm) that is not more than 0.08% Cs (at which $P_V = 104.932$ MW/m$^3$), and all of them well exceed our voluntary threshold criterion of 100 MW/m$^3$.

### 3.2. Potassium Seed and Air Oxidizer

The second set of results corresponds to the condition of using potassium vapor as the ionizable gas and using air as the oxidizer for the hydrogen combustion. Thus, compared to the previous set of results in the preceding subsection, the only change made here is changing the seed type from cesium to potassium.

Before presenting the results of this change, it is useful to mention that due to the higher ionization energy of potassium compared to cesium, potassium is harder to lose its valence electron in its outermost orbital ($4s^1$) [292–294] upon heating compared to cesium, whose outermost valence electron is in ($6s^1$) [295–297]. Therefore, it is expected that this change is accompanied by a drop in the H2MHD performance, primarily because of a decline in the electric conductivity. However, due to the nonlinearity of the problem and the present influence from other factors (particularly the speed of sound), the exact effect of this change in the seed type is not easy to predict analytically. Thus, our computational modeling becomes valuable in understanding quantitatively and qualitatively the implications of this single change.

In Table 9, we provide our computed thermochemical characteristics for the seeded gas mixture before ionization. Similar to the case of cesium seeding, the mixture's gas constant ($R_{mix}$) decreases (and the mixture's molecular weight increases) as the level of potassium increases. However, this decrease in the case of potassium additive is much weaker than in the case of cesium additive due to the weaker difference in the molecular weight between potassium (K) and either molecular nitrogen ($N_2$) or water vapor ($H_2O$). One mole of potassium (K) is only 1.396 times heavier than one mole of molecular nitrogen ($N_2$) and 2.170 times heavier than one mole of water ($H_2O$).

**Table 9.** Pre-ionization properties of potassium-seeded air–hydrogen combustion mixture.

| **Property** | **Different Pre-Ionization Potassium Mole Fraction ($X_K$)** | | | | |
|---|---|---|---|---|---|
| | **0.0625%** | **0.25%** | **1%** | **4%** | **16%** |
| $X_K$ [%] | 0.062500% | 0.250000% | 1.000000% | 4.000000% | 16.000000% |
| $X_{H2O}$ [%] | 34.688476% | 34.623395% | 34.363068% | 33.321763% | 29.156543% |
| $X_{N2}$ [%] | 65.249024% | 65.126605% | 64.636932% | 62.678237% | 54.843457% |
| $M_{mix}$ [kg/kmol] | 24.552 | 24.579 | 24.689 | 25.125 | 26.872 |
| $R_{mix}$ [J/(kg.K)] | 338.6451 | 338.2691 | 336.7734 | 330.9206 | 309.4112 |
| $\gamma_{mix}$ [-] | 1.243994 | 1.244278 | 1.245418 | 1.250087 | 1.270686 |
| $a$ [m/s] | 984.341 | 983.906 | 982.178 | 975.430 | 950.935 |
| $u$ [m/s] | 1968.68 | 1967.81 | 1964.36 | 1950.86 | 1901.87 |

For reference, the molecular weight of the combustion products without any seeding (34.710170% $H_2O$ and 65.289830% $N_2$ by mole) is 24.54304 kg/kmol; the unseeded mixture's gas constant is 338.7707 J/(kg.K); the unseeded mixture's specific heat ratio is 1.243900; and the unseeded mixture's speed of sound (at 2300 K) is 984.486 m/s.

Figure 4 shows how the electric conductivity of the hydrogen plasma monotonically and nonlinearly drops as the total pre-ionization absolute pressure increases. This is qualitatively similar to what was observed when cesium was the additive seed. Also, the existence of a peak electric conductivity at a certain level of potassium seed is similar to what was observed in the case of cesium. This potassium level that maximizes the electric conductivity is again (as was the case in the cesium case) approximately near 6%. It is noticeable that the electric conductivity (or any given level of seeding) dropped strongly when the seed is changed from cesium to potassium. For example, the maximum obtained electric conductivity in the case of cesium was 55.0610 S/m (at $X_{Cs} = 4\%$ and 0.0625 atm) is 2.55 times the maximum obtained electric conductivity in the case of potassium, which is 21.6201 S/m (at $X_K = 4\%$ and 0.0625 atm).

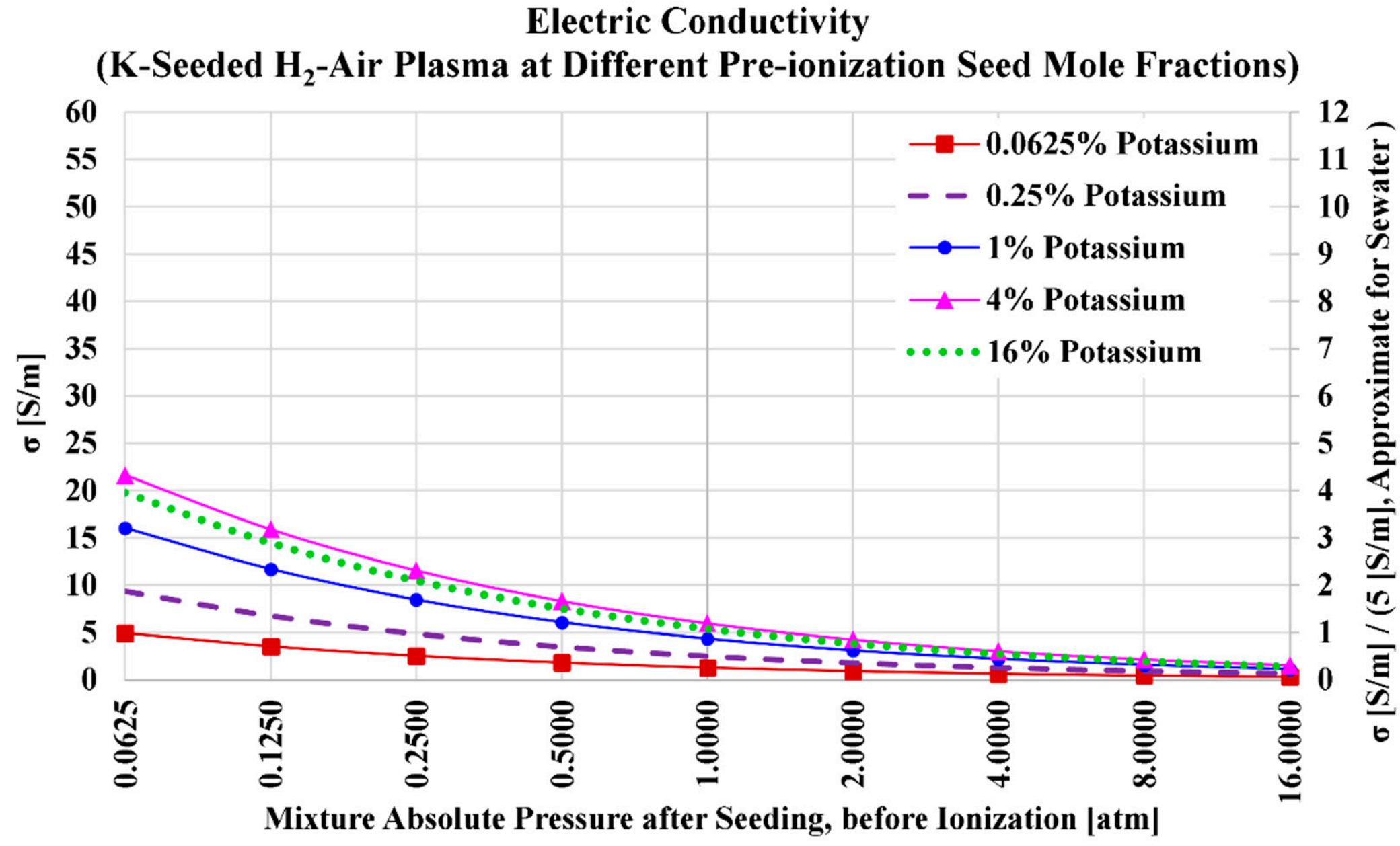

**Figure 4.** Electric conductivity in the case of potassium seed and air oxidizer.

Table 10 lists the numerical values of the plasma electric conductivity as visualized in the previous figure. We also add to the table the electric conductivity values at a pre-ionization cesium mole fraction of $X_K = 6\%$, which approximately corresponds to the maximized electric conductivity for this case (air–hydrogen combustion plasma with potassium seed).

**Table 10.** Electric conductivity [S/m] of hydrogen plasma (potassium seed, air oxidizer).

| Pressure [atm] | Plasma Electric Conductivity [S/m] (at the Different Potassium Levels) | | | | | |
|---|---|---|---|---|---|---|
| | **0.0625%** | **0.25%** | **1%** | **4%** | **6%** | **16%** |
| 0.0625 | 4.9217 | 9.3443 | 16.0343 | 21.6201 | 21.9827 | 19.7671 |
| 0.125 | 3.5282 | 6.7405 | 11.6895 | 15.8856 | 16.1504 | 14.4434 |
| 0.25 | 2.5189 | 4.8334 | 8.4459 | 11.5428 | 11.7342 | 10.4518 |
| 0.5 | 1.7931 | 3.4513 | 6.0630 | 8.3195 | 8.4568 | 7.5108 |
| 1 | 1.2739 | 2.4570 | 4.3325 | 5.9617 | 6.0597 | 5.3707 |
| 2 | 0.9037 | 1.7456 | 3.0859 | 4.2546 | 4.3243 | 3.8271 |
| 4 | 0.6405 | 1.2383 | 2.1931 | 3.0277 | 3.0772 | 2.7205 |
| 8 | 0.4536 | 0.8776 | 1.5561 | 2.1502 | 2.1853 | 1.9307 |
| 16 | 0.3211 | 0.6215 | 1.1029 | 1.5250 | 1.5499 | 1.3686 |

Considering the five selected seed levels (0.0625%, 0.25%, 1%, 4%, and 16%), the highest obtained electric conductivity is $\sigma = 21.6201$ S/m (at $X_K = 4\%$ and $p = 0.0625$ atm). At the higher $X_K = 16\%$ (and the same total pressure of 0.0625 atm), the electric conductivity is slightly lower, with a value of 19.7671 S/m; at the lower $X_{Cs} = 1\%$ (and the same total pressure of 0.0625 atm), the electric conductivity is mildly lower, with a value of 16.0343 S/m. If we take a total pressure of 1 atm (atmospheric MHD channel) as a reference, one can see that the electric conductivity is restricted to 6 S/m.

The ideal volumetric power densities ($P_V$) for the current H2MHD case of potassium seeding and air oxidizer are visualized in Figure 5 and listed in Table 11. We also add to the table the power density values at a pre-ionization cesium mole fraction of $X_K = 5\%$, which approximately corresponds to the maximized power density for this case (air–hydrogen combustion plasma with potassium seed).

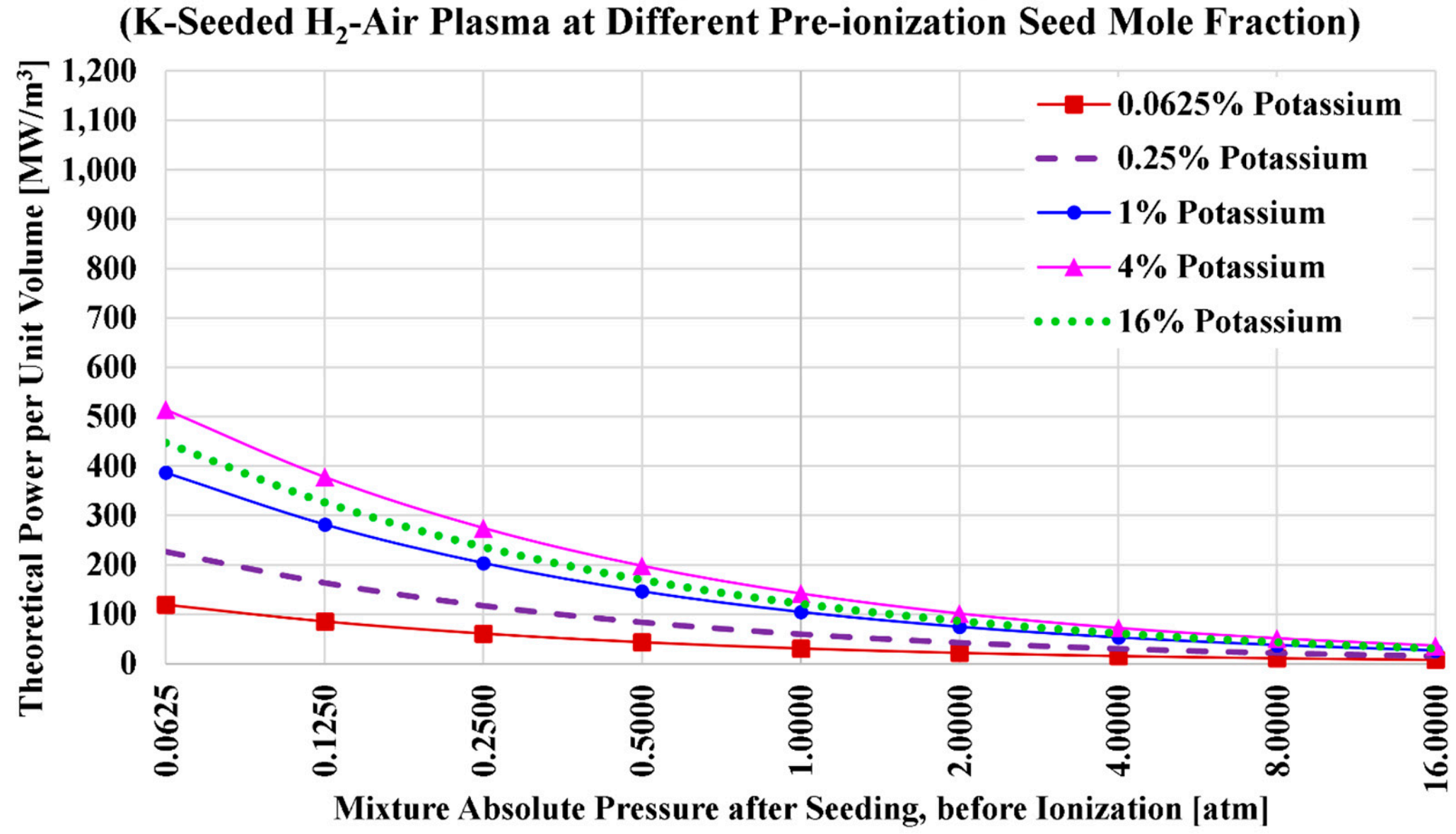

**Figure 5.** Ideal power density in the case of potassium seed and air oxidizer.

**Table 11.** Power density [MW/m$^3$] of hydrogen plasma (potassium seed and air oxidizer).

| Pressure [atm] | Volumetric Power Density [MW/m$^3$] (at the Different Potassium Levels) | | | | | |
|---|---|---|---|---|---|---|
| | **0.0625%** | **0.25%** | **1%** | **4%** | **5%** | **16%** |
| 0.0625 | 119.219 | 226.149 | 386.697 | 514.268 | 518.972 | 446.874 |
| 0.125 | 85.464 | 163.132 | 281.914 | 377.864 | 381.364 | 326.521 |
| 0.25 | 61.016 | 116.977 | 203.689 | 274.564 | 277.128 | 236.283 |
| 0.5 | 43.435 | 83.528 | 146.221 | 197.892 | 199.751 | 169.796 |
| 1 | 30.858 | 59.464 | 104.486 | 141.808 | 143.144 | 121.415 |
| 2 | 21.890 | 42.247 | 74.422 | 101.202 | 102.158 | 86.519 |
| 4 | 15.515 | 29.969 | 52.891 | 72.019 | 72.698 | 61.502 |
| 8 | 10.988 | 21.239 | 37.528 | 51.146 | 51.631 | 43.647 |
| 16 | 7.778 | 15.041 | 26.598 | 36.275 | 36.618 | 30.940 |

Unlike the case of cesium seeding, the peak power density with potassium seeding occurs near a seed level of 5% (not 3%), which is close to the potassium seeding level that maximizes the electric conductivity. This similarity of the potassium seed level ($X_K$) that maximizes the electric conductivity and the one that maximizes the power density is related to the near-neutral influence of the seed level on the speed of sound for the case of potassium seed (much weaker influence than the case of cesium seed). A nonlinear monotonic decline in the power density ($P_V$) as the total pressure ($p$) increases is noted for any potassium level, and this behavior resembles the one found earlier for the case of cesium seeding.

The power density with potassium seeding is clearly smaller than the one with cesium seeding (when other settings are kept the same). For example; considering the five selected seed levels (0.0625%, 0.25%, 1%, 4%, and 16%), the largest obtained power density at 1 atm in the case of potassium seeding is 141.808 MW/m$^3$ (with $X_K = 4\%$). This is 2.56 times the power density at 1 atm and $X_{Cs} = 4\%$, which was 362.806 MW/m$^3$. Although potassium seeding is associated with higher speeds of sound compared to cesium seeding, this small advantage for potassium seeding is largely counteracted by the big penalty in the electric conductivity due to changing the seed vapor from cesium to potassium.

Despite the drop in the power density in the case of potassium seeding, the power density can still exceed 100 MW/m$^3$ at 1 atm, with potassium levels as small as 1%. Thus, the H2MHD concept is still viable.

### *3.3. Potassium Seed and Oxygen Oxidizer*

The third set of results corresponds to the condition of using potassium vapor as the ionizable gas and using pure oxygen as the oxidizer for the hydrogen combustion. Thus, compared to the previous set of results in the preceding subsection, the only change made here is changing the oxidizer type from air (oxygen–nitrogen mixture) to pure oxygen.

Before presenting the results of this change, it is useful to mention that due to the absence of nitrogen from the combustion products, hydrogen combustion results in pure water vapor ($H_2O$). Water vapor has a relatively low molecular weight and thus has a relatively high specific gas constant. Therefore, the speed of sound in the current case of oxy-combustion of hydrogen is higher than the speed of sound in the previous case of air-combustion of hydrogen (for the same level of potassium seed). Therefore, it is expected that the change in the oxidizer (from air to oxygen) improves the speed of sound, and thus the plasma speed (given its fixed relation to the speed of sound in our study). It is thus the change in the electric conductivity that determines how eventually the power density is affected by the change in the oxidizer (from air to oxygen). This is difficult to predict without the numerical simulations performed in our study. This emphasizes the value and contribution made by our study to the field of MHD power generation through quantitatively and qualitatively investigating the influence of the oxidizer on the MHD power generation process.

In Table 12, we provide our computed thermochemical characteristics for the seeded gas mixture before ionization. As mentioned earlier, the specific gas constant and the speed of sound in the current case of oxy-hydrogen combustion are higher than their values in the previous case of air–hydrogen combustion. For example, with 1% potassium seed, the speed of sound with oxy-hydrogen combustion (1114.801 m/s) is 1.135 times the speed of sound with air–hydrogen combustion (982.178 m/s) with the same level of potassium seeding. Due to the larger molecular weight of potassium compared to water vapor, the more the added potassium, the higher the mixture's molecular weight and thus the lower the mixture's gas constant and the lower the speed of sound in that mixture of water vapor and potassium vapor. The mixture's specific gas constant ($\gamma_{mix}$) increases as the potassium level increases, but the decrease in the mixture's gas constant ($R_{mix}$) is stronger; thus, ultimately the speed of sound decreases when the potassium level increases.

Figure 6 shows the profiles of the electric conductivity of the hydrogen plasma. The monotonic and nonlinear drop as the total pre-ionization absolute pressure increases is retained, similar to the previous case of air–hydrogen combustion. However, there are two remarkable changes that can be observed from the figure when compared to its counterpart in the previous subsection that describes air–hydrogen combustion (with the same potassium seeding).

**Table 12.** Pre-ionization properties of potassium-seeded oxy-hydrogen combustion mixture.

| Property | Different Pre-Ionization Potassium Mole Fraction ($X_K$) | | | | |
|---|---|---|---|---|---|
| | **0.0625%** | **0.25%** | **1%** | **4%** | **16%** |
| $X_K$ [%] | 0.0625% | 0.25% | 1% | 4% | 16% |
| $X_{H2O}$ [%] | 99.9375% | 99.75% | 99% | 96% | 84% |
| $M_{mix}$ [kg/kmol] | 18.028 | 18.068 | 18.226 | 18.859 | 21.389 |
| $R_{mix}$ [J/(kg.K)] | 461.1850 | 460.1760 | 456.1837 | 440.8839 | 388.7337 |
| $\gamma_{mix}$ [-] | 1.183244 | 1.183490 | 1.184479 | 1.188543 | 1.206766 |
| $a$ [m/s] | 1120.311 | 1119.201 | 1114.801 | 1097.826 | 1038.727 |
| $u$ [m/s] | 2240.62 | 2238.40 | 2229.60 | 2195.65 | 2077.45 |

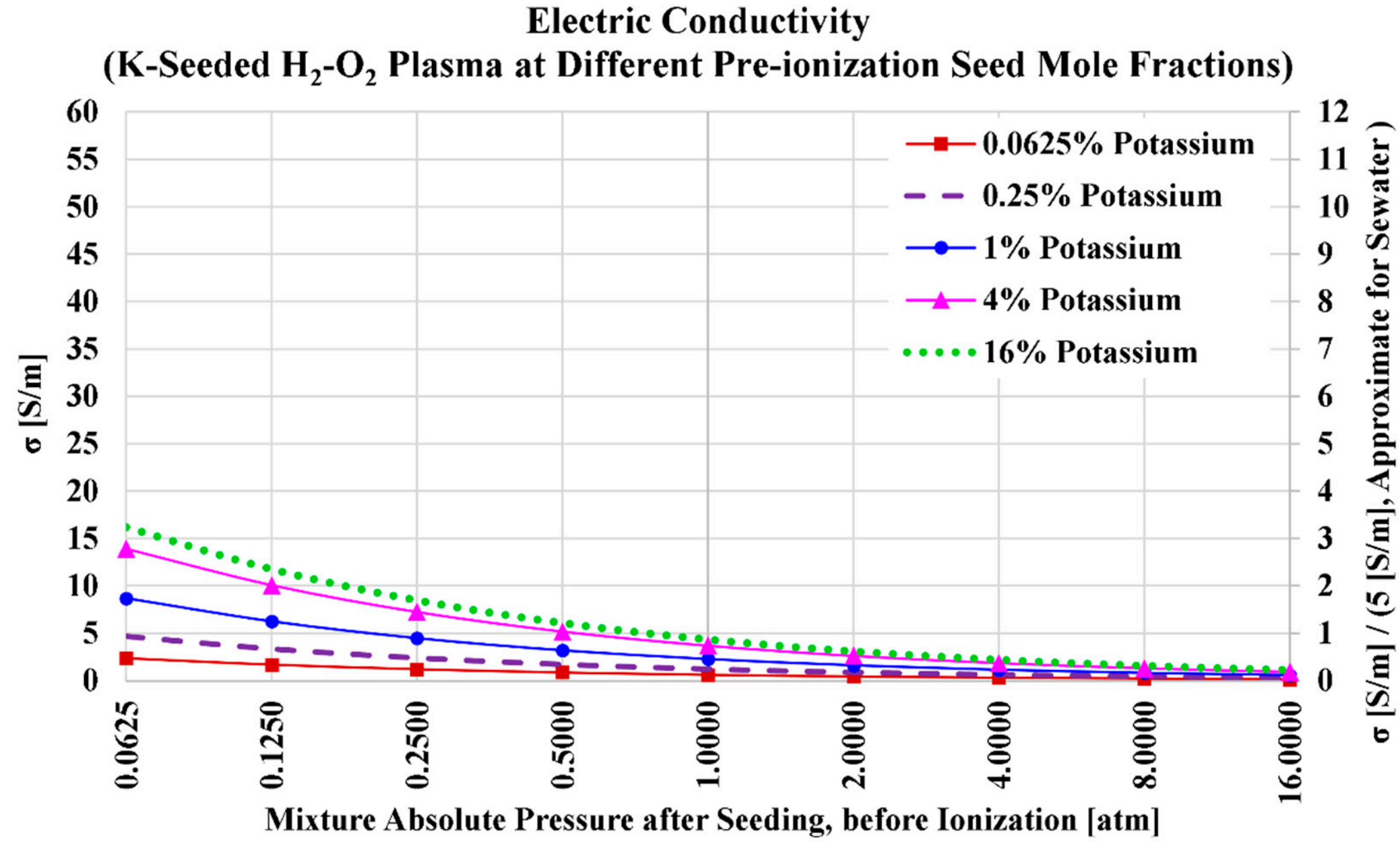

**Figure 6.** Electric conductivity in the case of potassium seed and oxygen oxidizer.

The first change worthy of attention is the overall suppression in the electric conductivity. This can be attributed to the larger effective electron-neutral momentum transfer collision cross-section for water vapor compared to molecular nitrogen [298–301]. Thus, when the molecular nitrogen in the combustion products is replaced with water vapor, electron mobility is retarded, and this suppresses the electric conductivity of the plasma.

The second change worthy of attention is that now with oxy-hydrogen combustion, the electric conductivity with 16% potassium is larger than the electric conductivity with 4% potassium. In the previous case of air–hydrogen combustion, the electric conductivity with 16% potassium was smaller than the electric conductivity with 4% potassium, due to the detrimental effect of over-seeding. In fact, while the electric conductivity with 16% potassium is larger than the electric conductivity with 4% potassium in the current case of oxy-hydrogen combustion, the peak electric conductivity has been already passed at $X_K = 16\%$. This means that this peaking phenomenon of the electric conductivity is actually still present in the current case of oxy-hydrogen plasma. The main difference is that this peak was easy to notice in the previous electric conductivity figure corresponding to air–hydrogen plasma with potassium seeding (in the previous subsection) because the potassium mole fraction ($X_K$) that maximized the electric conductivity was closer to the intermediate value of 4% than the terminal value of 16%. Changing the oxidizer from air to oxygen causes this particular ($X_K$) of maximized electric conductivity ($\sigma$) to shift to a larger position, becoming closer to the terminal value of 16% (but below it, located nearly at $X_K = 13\%$). Despite the gain in the electric conductivity when the potassium seed mole fraction ($X_K$) is increased from 4% to 16% (or 13%), the gain is too small to justify the need for exceeding the seed level beyond 4%.

Table 13 lists the numerical values of the plasma electric conductivity as visualized in the previous figure. We also add to the table the electric conductivity values at a pre-ionization potassium mole fraction of $X_K = 13\%$, which approximately corresponds to the maximized electric conductivity for this case (oxy-hydrogen combustion plasma with potassium seed).

**Table 13.** Electric conductivity [S/m] of hydrogen plasma (potassium seed, oxygen oxidizer).

| Pressure [atm] | Plasma Electric Conductivity [S/m] (at the Different Potassium Levels) | | | | | |
|---|---|---|---|---|---|---|
| | **0.0625%** | **0.25%** | **1%** | **4%** | **13%** | **16%** |
| 0.0625 | 2.3874 | 4.6735 | 8.6753 | 13.9154 | 16.2537 | 16.1854 |
| 0.125 | 1.7027 | 3.3408 | 6.2385 | 10.0857 | 11.8100 | 11.7558 |
| 0.25 | 1.2112 | 2.3803 | 4.4635 | 7.2563 | 8.5119 | 8.4704 |
| 0.5 | 0.8601 | 1.6920 | 3.1821 | 5.1932 | 6.0993 | 6.0683 |
| 1 | 0.6099 | 1.2008 | 2.2628 | 3.7030 | 4.3527 | 4.3300 |
| 2 | 0.4322 | 0.8513 | 1.6063 | 2.6335 | 3.0974 | 3.0809 |
| 4 | 0.3060 | 0.6030 | 1.1389 | 1.8696 | 2.1997 | 2.1878 |
| 8 | 0.2166 | 0.4269 | 0.8068 | 1.3256 | 1.5601 | 1.5516 |
| 16 | 0.1533 | 0.3021 | 0.5712 | 0.9391 | 1.1054 | 1.0993 |

Considering the five selected seed levels (0.0625%, 0.25%, 1%, 4%, and 16%), the highest obtained electric conductivity is $\sigma = 16.1854$ S/m (at $X_K = 16\%$ and $p = 0.0625$ atm). At the lower $X_K = 4\%$ (and the same total pressure of 0.0625 atm), the electric conductivity is mildly lower, with a value of 13.9154 S/m. If we take a total pressure of 1 atm as a reference, one can see that the electric conductivity is below 5 S/m (the approximate electric conductivity of seawater) regardless of the seeding level.

The ideal volumetric power densities ($P_V$) for the current H2MHD case of potassium seeding and oxygen oxidizer are visualized in Figure 7 and listed in Table 14. We also add to the table the power density values at a pre-ionization cesium mole fraction of $X_K = 9\%$, which approximately corresponds to the maximized power density for this case (oxy-hydrogen combustion plasma with potassium seed).

The electric conductivity ($\sigma$) at a potassium seed level ($X_K = 16\%$) was found to be larger than the electric conductivity at ($X_K = 4\%$). Similarly, the volumetric power density ($P_V$) at a potassium seed level ($X_K = 16\%$) is larger than the power density at ($X_K = 4\%$). However; because the speed of sound (and thus the plasma speed) at ($X_K = 16\%$) is larger than its value at ($X_K = 4\%$), the gap in ($P_V$) is diminished when the seed levels of 4% and 16% are compared.

When comparing the range of ($P_V$) here with those in the previous subsection (air-combustion instead of oxy-combustion); we notice that although the power density dropped, this drop is not large. For example, at 1 atm and 4% potassium, the power density here with oxy-combustion is 111.573 MW/m$^3$, while it was 141.808 MW/m$^3$ in the previous case of air-combustion. The difference (30.2350 MW/m$^3$) is 23.9% of the average of both values (126.6905 MW/m$^3$).

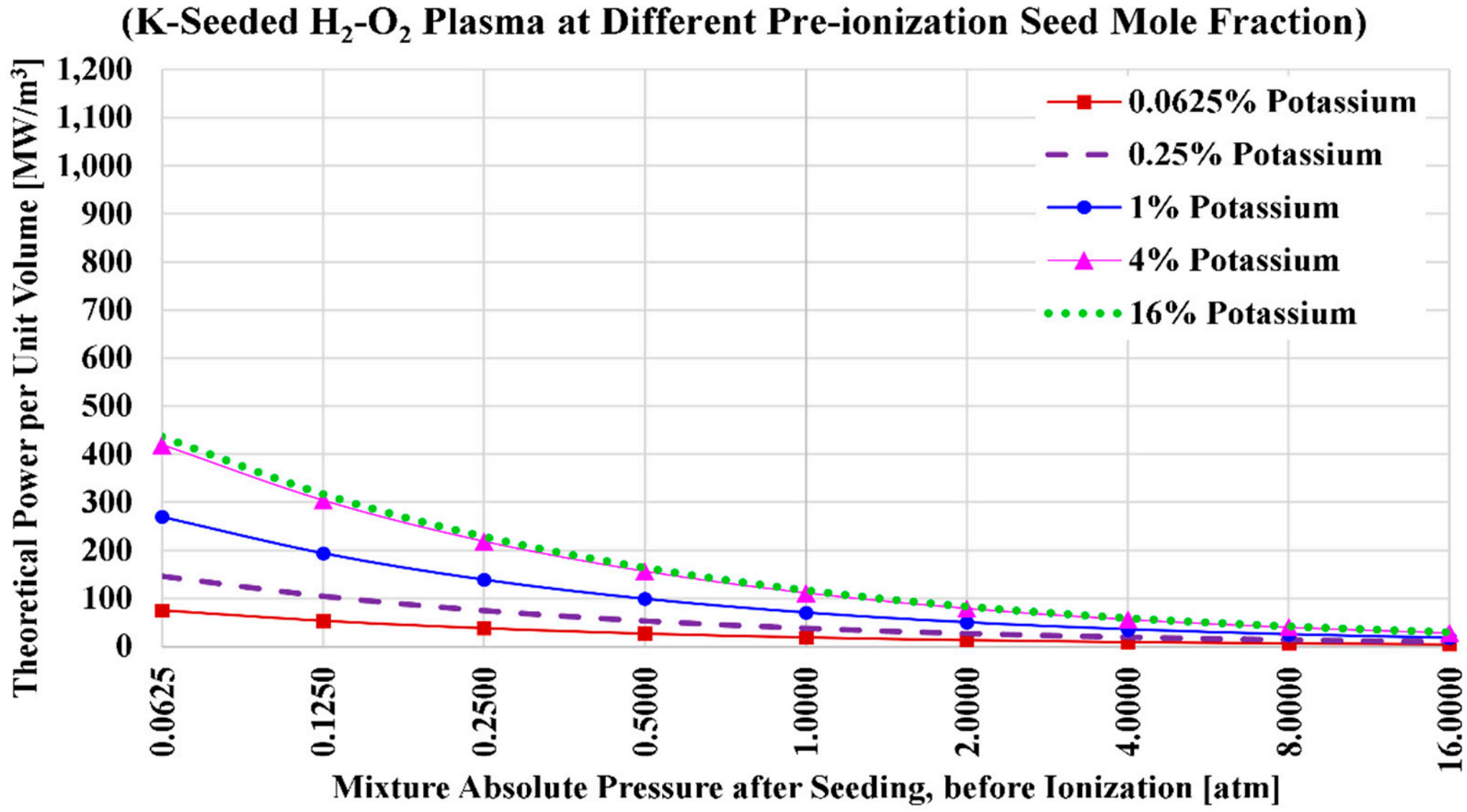

**Figure 7.** Ideal power density in the case of potassium seed and oxygen oxidizer.

**Table 14.** Power density [MW/m$^3$] of hydrogen plasma (potassium seed, oxygen oxidizer).

| Pressure [atm] | Volumetric Power Density [MW/m$^3$] (at the Different Potassium Levels) | | | | | |
|---|---|---|---|---|---|---|
| | **0.0625%** | **0.25%** | **1%** | **4%** | **9%** | **16%** |
| 0.0625 | 74.910 | 146.352 | 269.538 | 419.279 | 458.894 | 436.583 |
| 0.125 | 53.426 | 104.618 | 193.827 | 303.888 | 333.446 | 317.099 |
| 0.25 | 38.004 | 74.540 | 138.679 | 218.636 | 240.334 | 228.479 |
| 0.5 | 26.988 | 52.985 | 98.866 | 156.474 | 172.221 | 163.685 |
| 1 | 19.137 | 37.603 | 70.304 | 111.573 | 122.907 | 116.797 |
| 2 | 13.561 | 26.659 | 49.907 | 79.349 | 87.463 | 83.104 |
| 4 | 9.601 | 18.883 | 35.385 | 56.332 | 62.116 | 59.013 |
| 8 | 6.796 | 13.368 | 25.067 | 39.941 | 44.054 | 41.853 |
| 16 | 4.810 | 9.460 | 17.747 | 28.296 | 31.215 | 29.652 |

For exceeding the threshold power density of 100 MW/m$^3$, the potassium seed level should not be below 3% (at which we found that $a = 1103.376$ m/s and $\sigma = 3.4118$ S/m; thus, $P_V = 103.841$ MW/m$^3$)

### *3.4. Cesium Seed and Oxygen Oxidizer*

The fourth (and final) set of results corresponds to the condition of using cesium vapor as the ionizable gas and using pure oxygen as the oxidizer for the hydrogen combustion. Thus, compared to the previous set of results in the preceding subsection, the only change made here is changing the seed type from potassium to cesium.

In Table 15, we provide our computed thermochemical characteristics for the seeded gas mixture before ionization. The speed of sound drops faster with the cesium fraction compared to the previous case of potassium seeding, due to the large influence of cesium on the mixture's gas constant compared to potassium; this is because the cesium atom is 3.399 times heavier than the potassium atom. For example, at the same mole fraction of 4%, the speed of sound in the case of cesium seeding is 1002.552 m/s, which was larger (1097.826 m/s) in the case of potassium seeding. For comparison purposes, the speed of sound in pure water vapor (at 2300 K) is 1120.682 m/s. Thus, introducing a fraction of 4% cesium vapor causes a relative drop in this original speed of sound (1120.682 m/s) by 118.130 m/s or 10.54%. On the other hand, introducing a fraction of 4% potassium vapor causes a relative drop in the original speed of sound (1120.682 m/s) by only 22.856 m/s or 2.04%.

**Table 15.** Pre-ionization properties of cesium-seeded oxy-hydrogen combustion mixture.

| Property | Different Pre-Ionization Cesium Mole Fraction ($X_{Cs}$) | | | | |
|---|---|---|---|---|---|
| | **0.0625%** | **0.25%** | **1%** | **4%** | **16%** |
| $X_{Cs}$ [%] | 0.0625% | 0.25% | 1% | 4% | 16% |
| $X_{H2O}$ [%] | 99.9375% | 99.75% | 99% | 96% | 84% |
| $M_{mix}$ [kg/kmol] | 18.087 | 18.303 | 19.164 | 22.611 | 36.398 |
| $R_{mix}$ [J/(kg.K)] | 459.6900 | 454.2795 | 433.8539 | 367.7191 | 228.4336 |
| $\gamma_{mix}$ [-] | 1.183242 | 1.183482 | 1.184449 | 1.188419 | 1.206172 |
| $a$ [m/s] | 1118.493 | 1112.004 | 1087.161 | 1002.552 | 796.065 |
| $u$ [m/s] | 2236.99 | 2224.01 | 2174.32 | 2005.10 | 1592.13 |

In addition, it may be useful to add that for unseeded water vapor ($H_2O$), the specific gas constant is 461.5223 J/(kg.K) and the specific heat ratio is 1.183162.

Figure 8 shows how the electric conductivity of the hydrogen plasma varies with the total absolute pressure at the five selected cesium levels. The variations show a monotonical nonlinear drop, as observed in all previous cases (the three preceding subsections). However, compared to the previous subsection particularly, the electric conductivity is largely boosted here after replacing the seed from potassium (the previous subsection) to cesium (the current subsection). For example, at 1 atm and $X_{Cs} = 4\%$, the electric conductivity ($\sigma$) here is 11.0877 S/m. But in the previous case, at 1 atm and $X_K = 4\%$, this electric conductivity was 3.7030 S/m. The ratio of the two values of ($\sigma$) is 2.99.

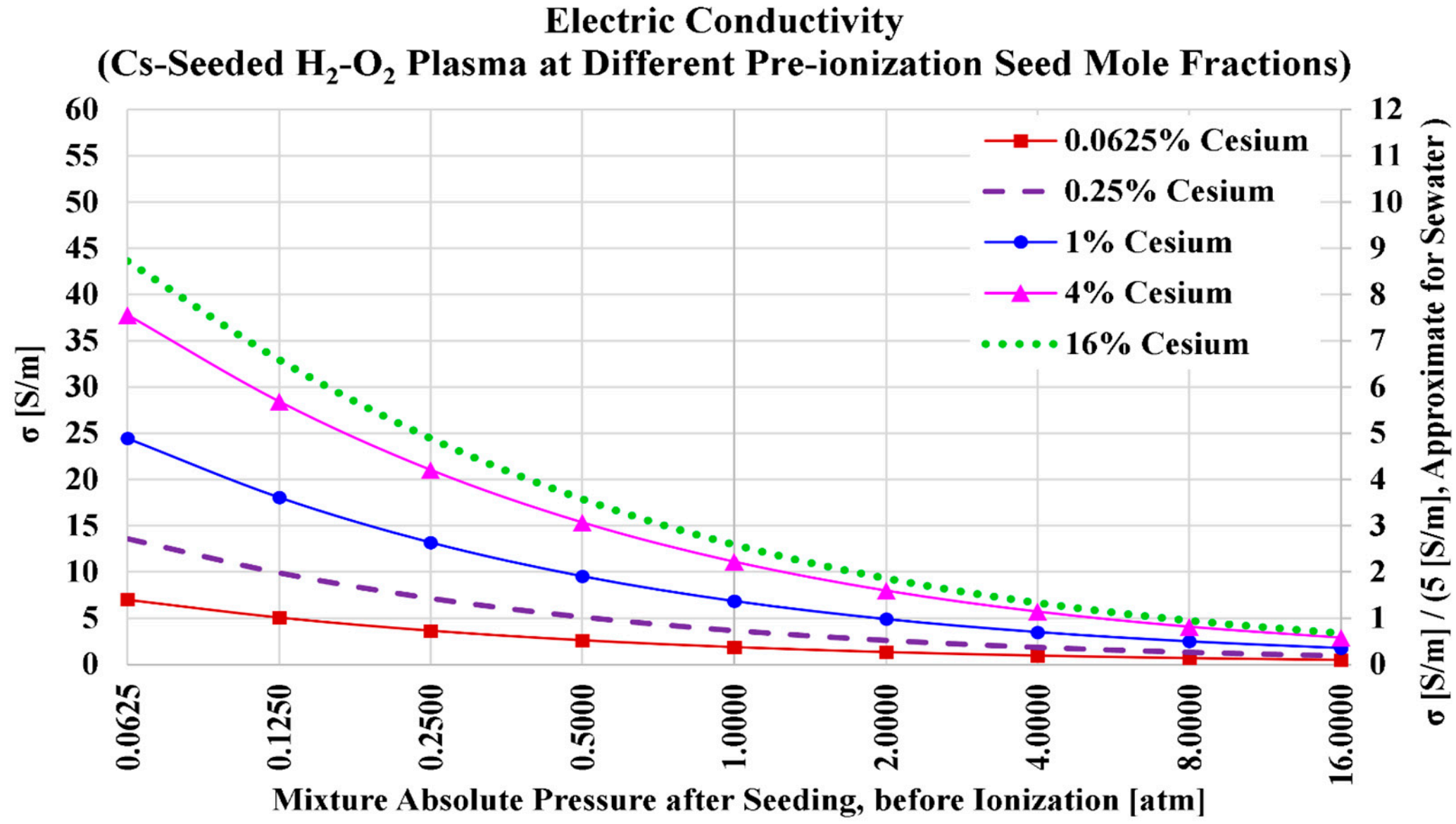

**Figure 8.** Electric conductivity in the case of cesium seed and oxygen oxidizer.

Also, as in the previous set of results (oxy-hydrogen combustion with potassium seeding), the seeding level at which the electric conductivity is maximized is larger than that level in the case of air–hydrogen combustion. That seeding level of maximum ($\sigma$) is approximately 13% in the current case of oxy-hydrogen combustion with cesium seeding (while it was near 6% in the case of air–hydrogen combustion with cesium seeding).

Table 16 lists the numerical values of the plasma electric conductivity as visualized in the previous figure. We also add to the table the electric conductivity values at a pre-ionization cesium mole fraction of $X_{Cs} = 13\%$, which approximately corresponds to the maximized electric conductivity for this case (oxy-hydrogen combustion plasma with cesium seed).

**Table 16.** Electric conductivity [S/m] of hydrogen plasma (cesium seed, oxygen oxidizer).

| Pressure [atm] | Plasma Electric Conductivity [S/m] (at the Different Cesium Levels) | | | | | |
|---|---|---|---|---|---|---|
| | **0.0625%** | **0.25%** | **1%** | **4%** | **13%** | **16%** |
| 0.0625 | 6.9878 | 13.5501 | 24.4282 | 37.7620 | 43.6710 | 43.5831 |
| 0.125 | 5.0633 | 9.8700 | 18.0514 | 28.4084 | 33.0271 | 32.9308 |
| 0.25 | 3.6427 | 7.1273 | 13.1716 | 21.0071 | 24.5184 | 24.4295 |
| 0.5 | 2.6073 | 5.1146 | 9.5223 | 15.3342 | 17.9480 | 17.8732 |
| 1 | 1.8595 | 3.6539 | 6.8382 | 11.0877 | 13.0034 | 12.9440 |
| 2 | 1.3227 | 2.6023 | 4.8876 | 7.9629 | 9.3514 | 9.3060 |
| 4 | 0.9392 | 1.8492 | 3.4818 | 5.6913 | 6.6899 | 6.6560 |
| 8 | 0.6661 | 1.3121 | 2.4746 | 4.0541 | 4.7684 | 4.7435 |
| 16 | 0.4719 | 0.9300 | 1.7559 | 2.8812 | 3.3901 | 3.3721 |

Considering the five selected seed levels (0.0625%, 0.25%, 1%, 4%, and 16%), the highest obtained electric conductivity is $\sigma = 43.5831$ S/m (at $X_{Cs} = 16\%$ and $p = 0.0625$ atm). At the lower $X_{Cs} = 4\%$ (and the same total pressure of 0.0625 atm), the electric conductivity is mildly lower, with a value of 37.7620 S/m (thus, 13.36% of the $\sigma$ value of 43.5831 S/m at 16% cesium is lost). If we take a total pressure of 1 atm as a reference, an electric conductivity of 10 S/m is achievable with a cesium mole fraction of 3% (at which the electric conductivity is 10.2333 S/m).

The ideal volumetric power densities ($P_V$) for the current H2MHD case of cesium seeding and oxygen oxidizer are visualized in Figure 9 and listed in Table 17. We also add to the table the power density values at a pre-ionization cesium mole fraction of $X_{Cs} = 5\%$, which approximately corresponds to the maximized power density for this case (oxy-hydrogen combustion plasma with cesium seed).

Like the case of cesium seeding with air–hydrogen combustion, but unlike the previous case of potassium seeding and oxy-hydrogen combustion, the power density here with 4% Cs is larger than the power density with 16% seed. This is a combined effect of how the electric conductivity ($\sigma$) and the speed of sound ($a$) respond to changes in the cesium mole fraction ($X_{Cs}$), which is different from their responses to changes in the potassium mole fraction ($X_K$).

The power density with cesium seeding (the current subsection) is clearly larger than the one with potassium seeding (the previous subsection).

A power density of 100 MW/m$^3$ or more is easily attainable at 1 atm, requiring a small mole fraction of cesium seed such as 0.25% only. Even with $X_{Cs} = 0.20\%$, we obtain $P_V = 101.838$ MW/m$^3$.

**Figure 9.** Ideal power density in the case of cesium seed and oxygen oxidizer.

**Table 17.** Power density [MW/m$^3$] of hydrogen plasma (cesium seed, oxygen oxidizer).

| Pressure [atm] | Volumetric Power Density [MW/m$^3$] (at the Different Cesium Levels) | | | | | |
|---|---|---|---|---|---|---|
| | **0.0625%** | **0.25%** | **1%** | **4%** | **5%** | **16%** |
| 0.0625 | 218.548 | 418.885 | 721.804 | 948.874 | 947.117 | 690.486 |
| 0.125 | 158.358 | 305.119 | 533.382 | 713.839 | 714.060 | 521.722 |
| 0.25 | 113.928 | 220.332 | 389.194 | 527.861 | 528.886 | 387.036 |
| 0.5 | 81.545 | 158.112 | 281.365 | 385.314 | 386.524 | 283.165 |
| 1 | 58.157 | 112.956 | 202.055 | 278.609 | 279.721 | 205.072 |
| 2 | 41.368 | 80.447 | 144.419 | 200.090 | 201.008 | 147.435 |
| 4 | 29.374 | 57.166 | 102.880 | 143.010 | 143.725 | 105.451 |
| 8 | 20.833 | 40.562 | 73.119 | 101.870 | 102.409 | 75.151 |
| 16 | 14.759 | 28.750 | 51.883 | 72.398 | 72.793 | 53.424 |

## 4. Discussion

In the current section, we discuss three topics that are related to the performed study but with a lower level of importance than the core results presented in the previous section. Therefore, skipping this discussion does not cause a deficiency in understanding the scope and outcomes of our study.

The first topic we discuss is a clarification of one of the study's findings that can be easily misinterpreted by the reader. This results-backed finding is that using air as an oxidizer is preferred to using oxygen in terms of the plasma electric conductivity and the output power. However, it should be noted that this preference for air as an oxidizer does not mean preferring an air-combustion process over an oxy-combustion process. The reason is that we assumed that the temperature remains the same when the oxidizer is changed from air to oxygen. However, oxy-combustion leads practically to elevated temperatures [302–304], which strongly boosts the electric plasma conductivity and thus the electric power output. Therefore, favoring air as an oxidizer is an outcome we found based on its influence on the chemical composition of the combustion products only, while factoring out its influence on the temperature of these combustion products. When the influence of the temperature due to oxy-fuel combustion is factored in, then the use of pure oxygen becomes preferred, because the large gain in the electric conductivity due to the increase in temperature is more important than the small decline in the electric conductivity due to the altered chemical composition of the plasma.

The second topic we discuss is the contribution made by our current study to the field of hydrogen magnetohydrodynamic power generation, in comparison to other studies in the literature. Although magnetohydrodynamic power generation is not a novel topic, our contribution is expressed through proposing the use of pure hydrogen as an environment-friendly fuel and quantifying the expected electric power under this proposed configuration over a wide range of settings. Due to the high nonlinearity of the problem's physics and the complicated relationships between various design variables and operating parameters [305–308], our study is viewed as a valuable support to the hydrogen magnetohydrodynamic power generation concept, where novel results are presented in detail as a preliminary design stage. In Table 18, we summarize the difference between our study and three recent studies (all dated 2025) in the same broad field of magnetohydrodynamic power generation. This emphasizes the distinction between our work and the work of others, and demonstrates the contributions claimed by our study.

**Table 18.** Comparison between our study and three recent studies in a similar field.

| Reference | Comparison with Our Study |
|---|---|
| [309] | The study in this reference is limited to potassium seeding and does not explore the effect of using cesium seeding (which we did). |
| [310] | The study in this reference uses liquid metal as a working fluid for magnetohydrodynamic power generators, not hydrogen (which we used). |
| [311] | The study in this reference describes a specific flow problem and a special channel that is porous. The exact fluid type is not specified, making the study have probable great theoretical value but not necessarily applicable to commercial magnetohydrodynamic power generation (which we considered). |

The third topic we discuss concerns the success of magnetohydrodynamic power generation compared to some other power technologies. Although magnetohydrodynamic power generation offers the advantage of no moving parts, direct power conversion, and high power density, this technology is not deployed commercially although the concept was proven experimentally [312–320]. Other power technologies (particularly solar photovoltaic modules and wind turbines) are successfully deployed as clean alternatives to traditional fossil-fuel power plants. The reasons for the delay in commercializing magnetohydrodynamic power generation can be attributed to the need for specialized powerful magnets, the high capital cost of construction, and the challenge in maintaining high electric conductivity [321–328]. Despite these challenges, research in magnetohydrodynamic power is still active [329–335], and new ideas may make the concept more competitive and attractive for realization.

## 5. Conclusions

In the current study, we reported novel results about our proposed concept of hydrogen magnetohydrodynamic power generation, particularly results about potential volumetric electric power densities under different operational conditions. We forecast the performance of a Mach 2 supersonic channel, operating with hydrogen-combustion plasma at a uniform temperature of 2300 K. The influence of the pressure, seed level, seed type, and oxidizer type was presented graphically and numerically.

The following findings can be stated:

Hydrogen magnetohydrodynamic generators permit concentrated power generation, with theoretical power densities as high as 300 MW/m$^3$ under a controlled condition of 1 atm and 1% cesium seeding (mole fraction), while using air as a conventional oxidizer.

Using cesium vapor as a seeding alkali metal for producing free electrons (as the main charge carriers in the plasma) is highly advantageous compared to potassium. The gain can more than double the output power.

Increasing the total static pressure monotonically decreases the electric conductivity and thus decreases the power density and performance of the H2MHD generator.

Increasing the alkali metal vapor seed amount monotonically increases the mixture's molecular weight, and thus monotonically decreases the mixture's specific gas density and in turn monotonically decreases the speed of sound in the mixture.

Increasing the alkali metal vapor seed amount increases the electric conductivity up to a certain level, where a peak electric conductivity is reached and then it declines with further seeding.

With air–hydrogen combustion and cesium seeding, the electric conductivity is maximized at a pre-ionization seeding mole fraction near 6%, while the power density is maximized (with a fixed Mach number) at a lower mole fraction near 3%.

With air–hydrogen combustion and potassium seeding, the electric conductivity is maximized at a pre-ionization seeding mole fraction near 6%, while the power density is maximized (with a fixed Mach number) at a lower mole fraction near 5%.

With oxy-hydrogen combustion and potassium seeding, the electric conductivity is maximized at a pre-ionization seeding mole fraction near 13%, while the power density is maximized (with a fixed Mach number) at a lower mole fraction near 9%.

With oxy-hydrogen combustion and cesium seeding, the electric conductivity is maximized at a pre-ionization seeding mole fraction near 13%, while the power density is maximized (with a fixed Mach number) at a lower mole fraction near 5%.

The future research direction of the current study may include (1) experimental investigation or a pilot-scale hydrogen magnetohydrodynamic generator (which was not possible for us to perform); (2) modeling the influence of plasma temperature (which was kept constant here); (3) modeling the influence of flue-gas recycle (FGR), where some of the hot steam leaving the generator is sent back to the generator and is mixed with fresh combustion products in order to exploit the high energy still contained within that exhaust steam; and (4) modeling the influence of replacing pure hydrogen as a fuel with either methane ($CH_4$) and a blend of hydrogen and methane. Because the Mach number and the applied magnetic-field flux were shown to have a simple quadratic correlation with the power density, they do not warrant a new separate study. However, the influence of the temperature is more complicated. It not only alters the electric conductivity in a nonlinear fashion, but also alters some of the thermodynamic properties.

**Supplementary Materials:** The following supporting information can be downloaded at: https://www.mdpi.com/article/10.3390/hydrogen6020031/s1.

**Funding:** This research received no external funding.

**Data Availability Statement:** Supplementary Data files related to the data reported in this research article are provided as four computer files, corresponding to the four main cases presented here (air–hydrogen combustion with cesium seeding, air–hydrogen combustion with potassium seeding, oxy-hydrogen combustion with potassium seeding, and oxy-hydrogen combustion with cesium seeding). Respectively, these data files are: airCs.csv, airK.csv, oxyK.csv, and oxyCs.csv.

**Conflicts of Interest:** The author declares that there are no known competing financial interests or personal relationships that could have appeared to influence the work reported in this paper.